\documentclass[10pt]{article}
\usepackage{mathtools, amssymb, mathrsfs, amsthm}
\usepackage[margin=1in]{geometry}
\usepackage{graphicx}
\graphicspath{{images/}}
\usepackage{caption}
\usepackage{subcaption}
\usepackage{enumitem}
\usepackage{titlesec}
\usepackage[utf8]{inputenc}
\usepackage[english]{babel}
\usepackage{miao}
\usepackage[font = small]{caption}
\usepackage[colorlinks = true, citecolor = blue, urlcolor = blue]{hyperref}
\usepackage[dvipsnames]{xcolor} 
\usepackage{cleveref}
\usepackage{autonum}
\usepackage{caption}
\usepackage{subcaption}
\usepackage{sectsty}
\usepackage{physics} 
\usepackage{comment}

\usepackage{algorithm}
\usepackage[noend]{algpseudocode}

\usepackage{parskip}
\newtheorem{lemma}{Lemma} 
\newtheorem{proposition}{Proposition}
\newtheorem{theorem}{Theorem}
\newtheorem{corollary}{Corollary}

\theoremstyle{definition}  
\newtheorem{remark}{Remark}
\newtheorem{example}{Example}

\newcommand{\myVar}{\operatorname{Var}}

\newcommand{\myCov}{\operatorname{Cov}}

\newcommand{\PSPA}{\textnormal{\texttt{PSPA}}}
\newcommand{\PPI}{\textnormal{\texttt{PPI}}}
\newcommand{\PPIpp}{\textnormal{\texttt{PPI++}}}

\def\A{{\bf A}}
\def\a{{\bf a}}
\def\B{{\bf B}}

\def\M{{\bf M}}

\def\S{{\bf S}}

\def\V{{\bf V}}

\def\X{{\bf X}}
\def\x{{\bf x}}
\def\Z{{\bf Z}}
\def\z{{\bf z}}

\def\calH{{\cal H}}
\def\calT{{\cal T}}

\def\calL{{\cal L}}
\def\calU{{\cal U}}

\def\bpsi{\boldsymbol\psi}

\def\bo{{\boldsymbol\omega}}

\def\bS{{\boldsymbol{\Sigma}}}
\def\bt{{\boldsymbol{\theta}}}
\def\bT{{\boldsymbol{\Theta}}}
\def\bta{{\boldsymbol{\eta}}}
\def\bphi{\boldsymbol\phi}

\def\inv{^{-1}}
\def\0{{\bf 0}}
\def\1{{\bf 1}}
\def\trans{^{\rm T}}
\def\star{^{*}}
\def\pr{\hbox{pr}}
\def\wh{\widehat}
\def\E{\mathbb{E}}

\def\log{{\rm log}}
\def\bit{\begin{itemize}}
\def\eit{\end{itemize}}
\def\meann{\dfrac{1}{n}\sum_{i=1}^n}
\def\meanN{\dfrac{1}{N}\sum_{i=n+1}^{n+N}}
\def\classical{_{\textnormal{C}}}
\def\tc{\wh\bt_{\textnormal{C}}}
\def\pp{_{\textnormal{PP}}}

\def\teff{\wh\bt_{\textnormal{EIF}}}

\def\teffs{\wh\bt_{\textnormal{EIF}\star}}
\def\effs{_{\textnormal{EIF}\star}}
\def\tprop{\wh\bt_{\PSPA}}
\def\prop{_{\textnormal{\PSPA}}}

\def\convp{\stackrel{P}{\rightarrow}}

\usepackage{tikz}
\usetikzlibrary{positioning}
\usepackage{authblk} 
\begin{document}
\title{Assumption-Lean and Data-Adaptive Post-Prediction Inference}
\author[1]{Jiacheng Miao}
\author[2]{Xinran Miao}
\author[1]{Yixuan Wu}
\author[1]{Jiwei Zhao}
\author[1]{Qiongshi Lu}

\affil[1]{Department of Biostatistics and Medical Informatics, University of Wisconsin-Madison}
\affil[2]{Department of Statistics, University of Wisconsin-Madison}
\maketitle

\begin{abstract}
A primary challenge facing modern scientific research is the limited availability of gold-standard data which can be costly, labor-intensive, or invasive to obtain. With the rapid development of machine learning (ML), scientists can now employ ML algorithms to predict gold-standard outcomes with variables that are easier to obtain. However, these predicted outcomes are often used directly in subsequent statistical analyses, ignoring imprecision and heterogeneity introduced by the prediction procedure. This will likely result in false positive findings and invalid scientific conclusions. In this work, we introduce PoSt-Prediction Adaptive inference (\PSPA) that allows valid and powerful inference based on ML-predicted data. Its “assumption-lean” property guarantees reliable statistical inference without assumptions on the ML prediction. Its “data-adaptive” feature guarantees an efficiency gain over existing methods, regardless of the accuracy of ML prediction. We demonstrate the statistical superiority and broad applicability of our method through simulations and real-data applications.
\end{abstract}

\section{Introduction}

Gene expression data obtained from various tissue and cell types can provide crucial insights into the coordinated biological mechanisms that drive disease etiology and characterize homeostasis \citep{lonsdale2013genotype}. However, some important tissues are often difficult to collect, which leads to underwhelming size of gene expression samples in those tissues. For example, the Genotype-Tissue Expression (GTEx) project is a comprehensive study focusing on gene expression regulation in many human tissues \citep{gtex2015genotype}. The percentage of individuals with missing gene expression ranges from 5\% to 90\% (median 47\%) across all tissues in GTEx \citep{gtex2020gtex}. This limits the scientific understanding of transcription regulation across tissue contexts.

The difficulty in obtaining gold-standard data certainly extends beyond human gene expression applications \citep{wang2023scientific}. While gold-standard data with high reliability are essential to the validity of scientific discoveries, obtaining them is often costly and labor-intensive. Fortunately, the advent and rapid development of machine learning (ML) have enabled prediction of outcomes using more accessible variables \citep{lecun2015deep}, showing great potential in reducing the need to acquire gold-standard data. 

Despite these benefits, replacing gold-standard data with ML predictions introduces new challenges, particularly in maintaining the validity of downstream statistical analyses. This issue is exemplified by the statistical analysis using imputed gene expression in the GTEx project. To address the insufficient sample size for rare tissues, several approaches have been proposed to impute gene expressions in these tissues using expression data from tissues that are easier to acquire \citep{wang2016imputing, vinas2023hypergraph, basu2021predicting}. However, imputed gene expression is often treated as if it were observed, and used directly in subsequent statistical analyses for scientific discovery \citep{vinas2023hypergraph}, such as exploring sex differences and the genetic basis of gene regulation. As demonstrated in \cite{wang2020methods, angelopoulos2023prediction}, direct use of ML prediction in statistical inference will most likely produce false positives findings, leading to invalid scientific conclusions.

\begin{figure}[H]
    \centering
    \includegraphics[width = 1\linewidth]{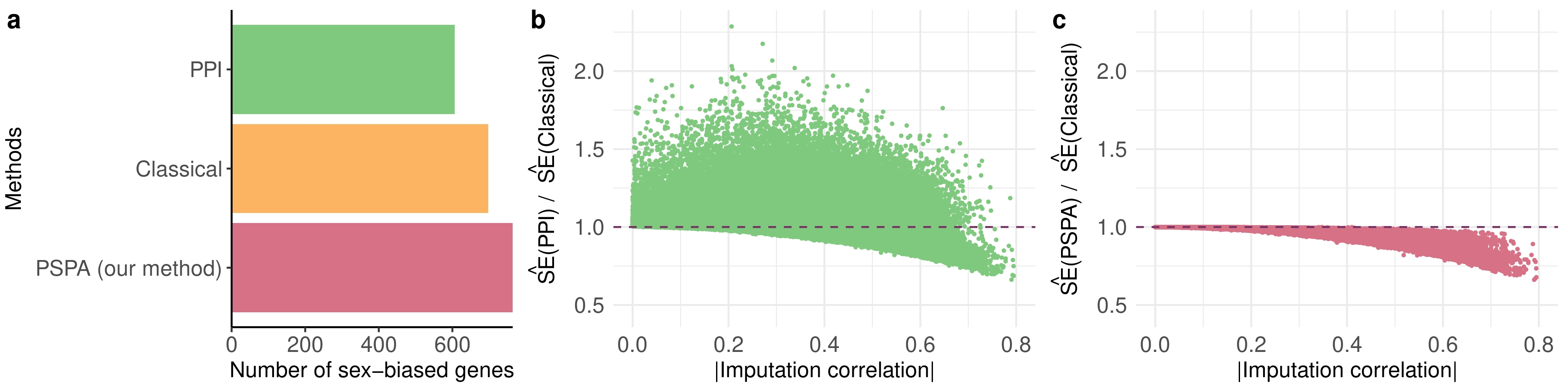}
        \caption{\textbf{Comparison of \PPI, \PSPA, and classical method in identifying sex-biased gene expressions using GTEx data.} 
        (a) number of sex-biased genes identified by each of the four approaches. (b) x-axis: absolute value of imputation correlation. y-axis: relative ratio of estimated standard error between \PPI~and classical method. (c) same as b but for our method \PSPA. The dashed line represents $y = 1$ in (b)-(c).}
    \label{fig:1}
\end{figure}

The challenge of making valid inferences from ML predictions extends beyond the specific example discussed, affecting numerous scientific disciplines where ML is applied \citep{bullock2020satellite}. Recent work has introduced a method called prediction-powered inference (\PPI), wwhich leverages a small set of labeled data with gold-standard outcomes and a large amount of unlabeled data with ML predictions to perform valid statistical inference \citep{angelopoulos2023prediction}. Although validity is guaranteed, \PPI~may be less efficient than the inference relying solely on labeled data. For instance, when applied to GTEx data, \PPI~ identifies fewer sex-biased gene expressions compared to the classical approach based on labeled data alone, highlighting a loss of statistical efficiency (Figure \ref{fig:1}). This limitation affects the broader applicability of \PPI~and similar methods.

To address this problem, we introduce a valid, powerful, and widely applicable approach named PoSt-Prediction Adaptive inference (\PSPA). \PSPA~is designed to integrate ML prediction with the observed gold-standard data to ensure valid and efficient statistical inference. We highlight two key features of our \PSPA~method:

\begin{itemize}
    \item \underline{Assumption-Lean}: The \PSPA~estimator is consistent and asymptotically normal, ensuring valid inference with no assumptions on the ML model and prediction accuracy. This means that \PSPA~can be used with arbitrarily misspecified ML prediction.
    \item \underline{Data-Adaptive}: The \PSPA~estimator is adaptive to the accuracy of the ML prediction. It utilizes more information from a ``good'' ML prediction to reduce variance and avoids variance inflation otherwise. Below, we will demonstrate that \PSPA~achieves element-wise variance reduction compared to existing methods, regardless of the ML prediction quality.
\end{itemize}

Our contributions are threefold: (i) We present a simple and data-adaptive method of post-prediction inference for an estimation problem defined through estimating equations, without any assumptions on the ML prediction or any parameterization of the true data-generating process. (ii) We establish the consistent and asymptotic normality of our proposed estimator. We further demonstrate the optimality of our estimator in terms of asymptotic variance over a class of estimators that includes existing estimators in post-prediction inference. (iii) In contrast to current methods which only focus on incorporating ML-predicted labels in statistical inference, our approach extends to utilizing both ML-predicted covariates and labels, enhancing the versatility and applicability of post-prediction inference.

The rest of this paper is organized as follows. In Section \ref{sec:problem}, we formulate our problem and illustrate our method with mean estimation to build an intuition. Next in Section \ref{sec:pspa}, we introduce a general protocol for applying \PSPA~to estimands defined through estimating equations. Further, we establish the asymptotic properties of the \PSPA~estimator and describe the estimation procedure. We build the connection with the semiparametric efficiency theory in Section \ref{sec:semi} and describe extensions in Section \ref{sec:extend}. Through extensive simulations (Section \ref{sec:simulation}) and real data analysis (Section \ref{sec:real}), we validate our theoretical claims and demonstrate the practical utility and effectiveness of the \PSPA~estimator. We then compared our method with a recent method \PPIpp~\citep{angelopoulos2023ppi++} in Section \ref{sec:ppi++}. This paper concludes with a discussion in Section \ref{sec:discussion}.

\section{Problem Formulation}\label{sec:problem}
Let $Y$ be a scalar outcome and $\X$ be $d$-dimensional covariates. The scientific interest is to estimate a $q$-dimensional parameter $\bt$ defined through an estimating equation,
\begin{align}\label{eq:esteq}
    \E\{\bpsi(Y,\X;\bt)\} = \0,
\end{align}
where $\bpsi(\cdot,\cdot;\bt)$ is a user-defined function. Such a definition of $\bt$ is very general including outcome mean, outcome quantile, least squares coefficients, or any other specific quantities of interest involving both $\x$ and $y$, including the unique minimizer of a loss function and the maximizer of a criterion function \citep{van2000asymptotic}.

We observe a random sample where only a small subset is labeled with outcome $Y$. In addition, we observe auxiliary variable $\Z$ that is predictive of the outcome $Y$.
Note that $\Z$ could be a subset of $\X$ and vice versa; see Remark \ref{remark:xz}.
Denote that sample as $\calL\cup\calU$ where $\calL = \{(y_i,\x_i,\z_i),i=1,\cdots,n\}$ and
$\calU=\{(\x_i,\z_i),i={n+1},\cdots,N+n\}$. Assume that $\calL$ and $\calU$ are independent and their marginal distributions of $(\X,\Z)$ are the same.
Let $n/N\to\rho$ as $n\to\infty$ and $N\to\infty$ and let $\rho = \infty$ if there is no unlabeled data.

We consider the availability of an external and independent prediction algorithm $\wh f(\cdot)$ on $\z$ that produces predictions $\wh f$ for the outcome variable $Y$. We assume that the operating characteristics of $\wh f$ are unknown to the user, and the data used to fit $\wh f$ are unavailable. $\wh f$ is considered a black box function and can be incorrect in predicting $y$.

With access to this black-box $\wh f$, our primary goal is to improve efficiency while making inference on the parameter $\bt$.
We refer to our problem as post-prediction inference, where inference occurs after predictions are made independently \citep{wang2020methods, angelopoulos2023prediction}. Specially, we make no assumptions on $\wh f$. The resulting protocol can benefit researches where training a prediction algorithm is unrealistic, and offers guidelines for scientists lacking computational resources or predictive modeling expertise.

\begin{example}[Sex-differentiated gene expressions]\label{example}
We present the scientific problem that motivates our setup: identifying sex-biased gene expressions in the brain cortex tissue using data from the GTEx project. For this analysis, we have labeled data denoted by $\calL = \{(y_i,\wh f_i, \x_i,\z_i),i=1,\cdots,205\}$, where $y_i$ represents the gene expression in brain cortex tissue, $\x_i$ includes a binary indicator for biological sex $in \{0, 1\}$, along with other technical factors such as surrogate variables for batch effects, age, RNA integrity number, and total ischemic time.  and total ischemic time. $\z_i$ is the whole-transcriptome gene expression profile in whole blood tissue that is easier to access and $\wh f$ is the imputed gene expression for in the brain cortex tissue using $\z_i$. In addition, our unlabeled data, $\calU=\{(\wh f_i, \x_i,\z_i),i={206},\cdots,670\}$, contains 465 samples whose gene expression was measured in whole blood but not in brain cortex. Our interest lies in the linear regression coefficient $\bt$ as a solution to $\E[\x(y - \x\trans\bt)] = 0$, focusing particularly on the coefficients corresponding to biological sex.
\end{example}

\begin{remark}[The consideration of $\X$ and $\Z$]\label{remark:xz} 
We split the covariate into two (possibly overlapping) parts, $\Z$ and $\X$, where the former is predictive of the outcome $Y$ and the latter appears in the estimating equation~\eqref{eq:esteq}. They could be the same, but a wide range of applications suggest they are different. When $\Z$ is not a subset of $\X$, the information it carries about the outcome $Y$ may be the most informative for the estimation problem~\eqref{eq:esteq}. Hence, a better prediction algorithm $\wh f$ does not necessarily correspond to a better estimate of $\bt$. 
As justified above, this work does not attempt to improve the prediction algorithm, but to empower the inference on the parameter of interest, given any arbitrary black-box $\wh f$.
\end{remark}

In the following, Section \ref{subsec:ex.mean} presents the key idea of the proposed method using mean estimation as an example. Section \ref{subsec:related.work} reviews related work. In Section \ref{sec:pspa}, we introduce the proposed framework for the estimation problem defined in equation~\eqref{eq:esteq}.


\subsection{Example: estimating the outcome mean}\label{subsec:ex.mean}
We provide the intuition of our approach using mean estimation. Consider $\psi(y,\x;\bt) = y - \theta$. The resulting estimand is $\theta = \E(Y)$, the outcome mean. Then the classical approach calculates an estimator from the sample average of observed outcomes, 
\begin{align}
    \wh\theta\classical=\meann y_i.
\end{align}

This estimator is unbiased, but can have high variance due to the small sample size of labeled data. To increase efficiency with the ML-predicted outcomes in unlabeled data, we propose to introduce another unbiased estimator of zero with a weighting scalar $\omega$,
\begin{align}
    \wh\theta\prop(\omega) = \meann y_i  + \omega\left[\dfrac{1}{N}\sum_{i=n+1}^{n+N} \wh f-\meann \wh f\right]
\end{align}
We refer to this estimator as the \PSPA~estimator. Both the classical and \PPI~estimator \citep{angelopoulos2023prediction} can be viewed as special cases of the \PSPA. The classical estimator is the \PSPA~estimator with $\omega=0$. The \texttt{PPI} estimator is the \PSPA~estimator with $\omega=1$:
\begin{align}
    \wh\theta\pp = \meann y_i +  \left[\dfrac{1}{N}\sum_{i=n+1}^{n+N}\wh f - \meann \wh f\right].
\end{align}

As detailed in Section \ref{sec:semi}, another candidate estimator is guided by the efficient influence function (EIF). We refer to this estimator as EIF$\star$-based where $\star$ represents the fact that the nuisance function is plugged in as an estimate,
\begin{align}
    \wh\theta\effs = \meann y_i +  \dfrac{N}{N+n}\left[\dfrac{1}{N}\sum_{i=n+1}^{n+N} \wh f - \meann \wh f\right].
\end{align}

This is a special case of \PSPA~estimator with $\omega=\frac{N}{N+n}$, achieving the semiparametric efficiency bound when $\wh f$ approximates its true counterpart at a sufficiently fast rate.

Since these estimators are summations of consistent estimators of the population mean and zero, they are consistent for the population mean. Additionally, Wald-type confidence intervals can be constructed using a consistent estimator for the asymptotic variance.

\begin{small}
\begin{align}\label{eq:var.mean}
\underbrace{
\frac{\myVar(Y)}{n}
+
\underbrace{
\left(
\omega^2\frac{\myVar[\wh f]}{n}
-
2\omega\frac{\myCov[Y,\wh f]}{n}
+
\omega^2\frac{\myVar[\wh f]}{N}
\right)}
_{\text {Additional terms}}
}_{\text {Variance of \PSPA~estimator}},
\end{align}
\end{small}

To guarantee that the asymptotic variance~\eqref{eq:var.mean} is no larger than the classical approach regardless of the quality of $\wh f$, the additional terms need to be no larger than zero. A key insight is that these additional terms are a quadratic function of $\omega$:
\begin{align}
    q(\omega) = \omega^2\left\{\frac{\myVar[\wh f]}{n}+\frac{\myVar[\wh f]}{N}\right\}-2\omega\frac{\myCov[Y, \wh f]}{n},
\end{align}

which achieves its minimum at $\omega^{\textnormal{opt}} = \myCov[Y, \wh f]/\{\myVar[\wh f] + n\myVar[\wh f]/N\}$ with minima
\begin{align}
q(\omega^{\textnormal{opt}}) = \dfrac{-\myCov[Y, \wh f]}{n\myVar[\wh f] + n^2\myVar[\wh f]/N} \leq 0,
\end{align}

where the equality holds if and only if $\myCov[Y, \wh f]=0$. In this case, our \PSPA~estimator is no less efficient than the classical estimator. In contrast, both \texttt{PPI} and EIF$\star$-based estimators can be less efficient than the classical estimator, when the ML predictions are not accurate.     
As we shall see in Section \ref{sec:pspa}, similar patterns appear in estimating the general parameter defined in (\ref{eq:esteq}). 
In addition, the additional terms are below zero if and only if $\myCov[Y, \wh f] > \frac{\omega (N+n)\myVar[\wh f]}{2N}$. This indicates that in order for \PPI~to achieve efficiency improvement
(where $\omega = 1$), the prediction accuracy of the ML algorithm cannot be low.

\begin{remark}[Data-Adaptive Feature]
    We provide an intuition for the ``data-adaptive'' feature of \PSPA. When the ML algorithm is purely random, $\myCov[Y, \wh f]=0$ and hence the optimal weight $\omega^{\textnormal{opt}} = 0$. Therefore, $\PSPA$~degenerates to the classical estimator that uses only labeled data. This suits the expectation that such ML prediction should not be used in inference. On the other hand, if the ML prediction is perfect, $\myCov[Y, \wh f] = \myVar[\wh f] = \myVar[Y]$ and hence the optimal weight is $\omega^{\textnormal{opt}} = N/(N+n)$. In this case, $\PSPA$~degenerates to the EIF*-based estimator that achieves the semiparametric efficiency bound. This is also intuitive since perfect ML prediction can be treated as gold-standard data and the estimator should utilize all the (measured and predicted) information and thus is efficient.
\end{remark}


\subsection{Related work}
\label{subsec:related.work}
Our setting is closely related to a recent method called prediction-powered inference (\PPI) \citep{angelopoulos2023prediction}, which ensures valid inference with gold-standard labels and arbitrary ML predictions. It corresponds to a special case of our \PSPA~estimator where $\omega = 1$ (or a vector of all ones). Both theoretical and numeric comparisons suggest that the \PPI~estimator is less efficient than the proposed \PSPA~estimator. Our approach is also closely related to a concurrent method called \PPIpp~\citep{angelopoulos2023ppi++}, which also aims to enhance the efficiency of \PPI. \PSPA~is guaranteed to be more efficient than \PPIpp. Unlike \PSPA, \PPIpp~cannot guarantee element-wise variance reduction compared with the classical method in multi-dimensional estimation problems. Additionally, \PPIpp~cannot be applied to statistical inference with ML-predicted covariates, while \PSPA~is capable of handling this type of applications. We present a comprehensive comparison of \PSPA~and \PPIpp~in Section \ref{sec:ppi++}.

Our proposal of boosting a consistent estimator with a consistent estimator of zero has been a long-standing idea. One famous estimator is the augmented inverse probability weighting estimator in the literature of missing data \citep{robins1994estimation} and causal inference \citep{robins2000robust}, in which a consistent inverse probability weighting estimator is augmented by a weighted residual term with mean zero, resulting in efficiency gain \citep{hahn1998role}. In fact, it reaches the semiparametric efficiency bound \citep{bickel1993efficient,hahn1998role} when nuisance functions are correctly specified or approximated sufficiently fast. Although the pursuit of efficiency under correct specifications differs from our main focus, we establish the efficiency bound of our estimation problem~\eqref{eq:esteq} in Section \ref{sec:semi} and illustrate the connection between the \PSPA~estimator and the efficient influence function.

The idea of using unlabeled data fits into the broader context of improving efficiency with auxiliary data, which has been widely applied in survey estimation \citep{breidt2017model}, missing data \citep{robins1994estimation}, measurement error models \citep{chen2005measurement}, causal inference with surrogate outcomes \citep{kallus2020role}, and semi-supervised learning \citep{wang2007large,chakrabortty2018efficient}. The literature on semi-supervised learning has adopted a similar data structure to our problem in which labeled data are accompanied by unlabeled data (possibly with surrogate outcomes). 
Recently, \citet{chakrabortty2018efficient} have considered efficiency improvement in linear models  under possible model misspecifications. After that, estimation methods have been proposed for mean estimation \citep{zhang2019semi}, best linear predictor estimation \citep{azriel2022semi}, and general M-estimation problems \citep{song2023general}. 
Our work is distinct from this line of research in that we consider post-prediction inference with any ``black-box'' ML predictions for efficiency improvement.

\section{\PSPA~for assumption-lean and data-adaptive post-prediction inference }\label{sec:pspa}
In this section, we introduce our method \PSPA~for assumption-lean and data-adaptive post-prediction inference. For estimand defined by~\eqref{eq:esteq} and any ML prediction, it guarantees the validity of inference results and element-wise variance reduction compared with the classical method that uses labeled data alone. We present our estimator and algorithm in Section \ref{subsec:alg} and establish the theoretical guarantees for our method in Section \ref{subsec:theory}. Examples of applying our method for regression tasks are provided in Section \ref{subsec:ex.regression}.

\subsection{\PSPA~estimator}\label{subsec:alg}
Bearing in mind that $\wh f$ could be incorrect, one typical approach to estimating $\bt$ is to ignore the unlabeled data and use only the labeled data. This classical estimator $\tc$ solves for 
\begin{align}
    \meann \bpsi(y_i,\x_i;\bt) = \0. 
\end{align}
It is always consistent, but may have low efficiency because it ignores the unlabeled data.

In this work, we aim to provide an estimator that is consistent and more efficient than the classical estimator, regardless of the quality of the $\wh f$.
Similar to the mean estimation example, we add an augmentation term, which has a mean of zero and is indexed by a vector $\bo=[\omega_1,\cdots,\omega_q]\trans$, to the estimation equation for the classical approach:
\begin{align}\label{eq:pspa.finite.ee}
  \diag(\bo)\left\{-\meann\bpsi(\wh f,\x_i;\bt) + \dfrac{1}{N}\sum_{i=n+1}^{n+N} \bpsi(\wh f,\x_i;\bt)\right\},
\end{align}
where $\diag(\bo)$ is a diagonal matrix with diagonal elements $\omega_1$, $\cdots$, $\omega_q$. We refer to the vector $\bo$ as a weighting vector as it controls the contribution of labeled outcomes and the arbitrary prediction $\wh f$ in a data-adaptive way. Together, we propose our \PSPA~estimator $\wh\bt\prop({\bo})$ that solves the equation,
\begin{align}\label{eq:pspa.estimator}
\Psi_{\PSPA}^\bo(\bt) := \meann\bpsi(y_i,\x_i;\bt) + \diag(\bo)\left\{-\meann\bpsi(\wh f,\x_i;\bt) + \dfrac{1}{N}\sum_{i=n+1}^{n+N} \bpsi(\wh f,\x_i;\bt)\right\} = \0,
\end{align}

for a given weighting vector $\bo$. This equation corresponds to a class of estimators $\wh\bt\prop({\bo})$ for different fixed values of $\bo$. We next establish the consistency and asymptotic normality of these estimators with the following regularity conditions. Condition (C1) is a regularity assumption on the parameter of interest. Conditions (C2)-(C4) guarantee consistency. Condition (C5) is needed for asymptotic normality. All conditions are reasonable and standard \citep{van2000asymptotic}.

\begin{itemize}
    \item[(C1)] The parameter space $\bT$ is a compact subset of $\mathbb{R}^q$, containing the true parameter $\bt_0$ as a unique solution to~\eqref{eq:esteq}.
    \item[(C2)] The function in~\eqref{eq:pspa.estimator} is continuous and the equation has a unique solution.
    \item[(C3)]  $\sup_{\bt \in \Theta}\E[\|\bpsi(Y,\X;\bt)\|^2] < \infty$
    and
    $ \sup_{\bt \in \Theta}
    \E[\|\diag(\bo)\cdot\bpsi(\wh f(\Z),\X;\bt)\|^2] < \infty$ where $\|\cdot\|$ is the $L_2$-norm that takes over the true data-generating process.
    \item[(C4)]  Labeled and unlabeled data are i.i.d drawn from the population of interest.
    \item[(C5)] There exists a function $\boldsymbol{\varphi}$ such that $\E[\sup_{\bt \in \Theta}\boldsymbol{\varphi}^{2}] < \infty$ and $\E \lVert\bpsi(Y,\X;\bt') - \bpsi(Y,\X;\bt'')\lVert_2^2\leq \boldsymbol{\varphi} \lVert\bt'-\bt''\rVert^2$ for any $\bt'$ and $\bt''$ in a neighborhood of $\bt$. Furthermore, $\E\{\bpsi(y,\x;\bt)\}$ is differentiable at $\bt$ with a nonsingular derivative matrix.
\end{itemize}

\begin{theorem}\label{thm:asy.normal.pspa} 
Under Conditions (C1)-(C4), assuming $\frac{n}{N} \rightarrow \rho$ as $n\to\infty$ and $N \rightarrow \infty$, then the proposed estimator $\wh\bt\prop({\bo})$ converges to $\bt$ in probability. Assuming additionally Condition (C5), we have
\begin{align}\label{eq:asy.normal}
    \sqrt n(\wh\bt\prop({\bo}) - \bt)\coD \mathcal{N}(\0,\bS(\bo)),
\end{align}
where $\bS(\bo) = \A\inv\V(\bo)\A\inv$, $\A = \E[\partial\bpsi(Y,\X;\bt)/\partial\bt]$,
$\V(\bo) = \M_1 + \diag(\bo)(\M_2 + \rho\M_3)\diag(\bo) - 2\diag(\bo)\M_4$,
$\M_1 = \myVar[\bpsi(Y,\X;\bt)]$, $\M_2 = \myVar[\bpsi(\wh f(\Z),\X;\bt)]$, $\M_3 = \myVar[\bpsi(\wh f(\Z),\X;\bt_0)]$, \\$\M_4 = \myCov[\bpsi(Y,\X;\bt),\bpsi(\wh f(\Z),\X;\bt)]$. 
\end{theorem}

The proof is contained in Appendix \ref{proof:thm.asy.normal}. Theorem \ref{thm:asy.normal.pspa} suggests that the proposed estimator $\tprop({\bo})$ is asymptotically normal with a fixed weighting vector $\bo$. In practice, this weighting vector $\bo$ should be estimated from data with the pursuit of efficiency. We provide an estimation procedure aimed at element-wise variance  reduction. This is made possible by a key observation that the $j$-th diagonal element of the asymptotic covariance matrix $\bS(\bo)$ is a quadratic function of $\omega_j$ and does not depend on other components of $\bo$:
\begin{align}
    \bS_{jj}(\bo) 
    =
    \omega_j^2[\A\inv(\M_2+\rho\M_3)\A\inv]_{jj} - 2\omega_j[\A\inv\M_4\A\inv]_{jj} + [\A\inv\M_1\A\inv]_{jj}
\end{align}

where $[\M]_{jj}$ represents the $j$-th diagonal element of matrix $\M$. We define the optimal weighting vector $\bo^{\textnormal{opt}} = [\omega_1^{\textnormal{opt}},\dots,\omega_q^{\textnormal{opt}}]\trans$ such that each coordinate of the weighting vector minimizes the asymptotic variance for the corresponding coordinate:
\begin{align}\label{eq:omega.opt.explicit}
    \omega_j^{\textnormal{opt}} &= \arg\min_{\omega_j}\bS_{jj}(\omega_j)=  \dfrac{[\A\inv\M_4\A\inv]_{jj}}{[\A\inv(\M_2+\rho\M_3)\A\inv]_{jj}} \text{ for all 
 } j \in 1, \dots, q,
\end{align}

The resulting asymptotic variance for $j$-th coordinates is
\begin{align}
    \bS_{jj}(\bo^{\textnormal{opt}}) &=  [\A\inv\M_1\A\inv]_{jj} - 
    \dfrac{[\A\inv\M_4\A\inv]_{jj}}{ [\A\inv(\M_2+\rho\M_3)\A\inv]_{jj}}
    \label{eq:var.j.opt}.
\end{align}

Such construction of $\bo^{\textnormal{opt}}$ guarantees the optimality of the element-wise asymptotic variance across the class of estimators with any weighting vector $\bo$.  Next, we present our algorithm for estimation and inference.

\begin{algorithm}[H]
  \caption{\PSPA~estimation with ML-predicted labels}
  \label{alg:pspa:y}
  \begin{algorithmic}[1]
  \Require Data $\calL\cup\calU$, pre-trained ML model $\wh f$, error rate $\alpha \in (0, 1)$.
  \State Obtain the classical estimator $\wh\bt_{\text{C}}$ by solving $\meann\bpsi(y_i,\x_i;\bt) = \0$.
  \State Obtain the optimal weighting vector $\wh\bo^{\textnormal{opt}} = [\wh\omega_1^{\textnormal{opt}},\dots,\wh\omega_q^{\textnormal{opt}}]\trans$ by
  $$
  \wh\omega_j^{\textnormal{opt}} =  \min(\frac{[\wh\A_{\text{C}}\inv\wh\M_{4,{\text{C}}}\wh\A_{\text{C}}\inv]_{jj}}{[\wh\A_{\text{C}}\inv(\wh\M_{2,{\text{C}}}+\rho\wh\M_{3,{\text{C}}})\wh\A_{\text{C}}\inv]_{jj}}, 1) \text{ for all 
 } j \in 1, \dots, q,
  $$
where $\wh\A_{\text{C}}$, $\wh\M_{1,{\text{C}}}$, $\wh\M_{2,{\text{C}}}$, $\wh\M_{3,{\text{C}}}$, $\wh\M_{4,{\text{C}}}$ are sample analogs of $\A$, $\M_1$, $\M_2$, $\M_3$, $\M_4$ with $\wh\bt_{\text{C}}$ plugged in.
\State Obtain the \PSPA~estimator with a one-step update:
   $$ \wh\bt_{{\PSPA}} = \wh\bt_{\text{C}} - \left[\nabla\Psi_{\PSPA}^{\wh\bo^{\textnormal{opt}}}(\wh\bt_{\text{C}})\right]\inv
       \Psi_{\PSPA}^{\wh\bo^{\textnormal{opt}}}(\wh\bt_{\text{C}}),
       \text{ where } \nabla\Psi_{\PSPA}^{\wh\bo^{\textnormal{opt}}} = \partial\Psi_{\PSPA}^{\wh\bo^{\textnormal{opt}}}(\bt)/\partial\bt\vert_{\bt = \wh\bt_{\text{C}}} $$

\State Obtain the asymptotic variance of $\wh\bt_{{\PSPA}}$:
$$
    \wh\bS(\wh\bo^{\textnormal{opt}}) = \wh\A_{\PSPA}\inv[\wh\M_{1,\PSPA} + \diag(\wh\bo^{\textnormal{opt}})(\wh\M_{2,\PSPA} + \rho\wh\M_{3,\PSPA})\diag(\bo) - 2\diag(\wh\bo^{\textnormal{opt}})\wh\M_{4,\PSPA}]\wh\A_{\PSPA}\inv,
$$
where $\wh\A_\PSPA$, $\wh\M_{1,\PSPA}$, $\wh\M_{2,\PSPA}$, $\wh\M_{3,\PSPA}$, $\wh\M_{4,\PSPA}$ are sample analogs of $\A$, $\M_1$, $\M_2$, $\M_3$, $\M_4$ with $\wh\bt_\PSPA$ plugged in.

  \Ensure \PSPA~estimator $\wh\bt_{{\PSPA}}$, standard error  $\sqrt{\frac{\wh\bS(\wh\bo^{\textnormal{opt}})_{jj}}{n}}$, $\alpha$-level confidence interval $\mathcal{C}_{\alpha,j}^{\textnormal{\PSPA}}=(\wh\bt_{\textnormal{\PSPA}_j} \pm z_{1-\alpha / 2} \sqrt{\frac{\wh\bS(\wh\bo^{\textnormal{opt}})_{jj}}{n}})$, and (two-sided) p-value $ 2 ( 1 - \Phi(| \frac{\wh{\bt}_{\PSPA_j}}{\sqrt{\frac{\wh\bS(\wh\bo^{\textnormal{opt}})_{jj}}{n}})} | ) )$ for the $j$-th coordinate. Here, $\Phi$ is the CDF of the standard normal distribution.
   \end{algorithmic}
\end{algorithm}

\subsection{Theoretical guarantees}\label{subsec:theory}
Next, we establish the theoretical guarantees for our proposed estimator and algorithm. We modify the Theorem \ref{thm:asy.normal.pspa} to reflect that $\wh\bo$ in the algorithm is estimated from the data.

\begin{corollary}\label{cor:asy.normal.est.w}
    Suppose $\wh\bo \coP \bo$ and conditions for Theorem \ref{thm:asy.normal.pspa} hold, then $\wh\bt\prop({\bo}) \coP \bt$ and
    \begin{align}
        \sqrt{n}(\wh\bt\prop({\bo})-\bt)\coD N(\0,\bS(\bo)).
    \end{align}
\end{corollary}

Proof of Corollary \ref{cor:asy.normal.est.w} is contained in Appendix \ref{proof:asy.normal.est.w}. In our algorithm, we substitute the sample analogs for $\A$, $\M_1$, $\M_2$, $\M_3$, $\M_4$ into $\omega_j^{\textnormal{opt}}$. Given that these sample analogs are typically consistent estimator, Slutsky's theorem implies that $\wh\bo^{\textnormal{opt}} \coP \bo^{\textnormal{opt}}$. Therefore, this additional condition is satisfied by our algorithm. The asymptotic normality in Corollary \ref{cor:asy.normal.est.w} guarantees the validity for inference for our algorithm by the following Corollary \ref{cor:ci}.

\begin{corollary}\label{cor:ci}
Suppose $\wh\bS(\wh\bo) \coP \bS(\bo)$ and conditions for Theorem \ref{thm:asy.normal.pspa} hold. Given confidence level $1-\alpha\in(0,1)$,
   \begin{align}
    \lim_{n \rightarrow \infty}\mathbb{P}(\theta_{j}\in\mathcal{C}_{\alpha,j}^{\wh\bo})\geq 1-\alpha,\ j=1\cdots,q,
  \end{align}
  
  where $\theta_{j}$ is the $j$-th coordinate of the parameter and  $\mathcal{C}_{\alpha,j}^{\textnormal{\PSPA}}=(\wh\bt_{\textnormal{\PSPA}_j}(\wh\bo) \pm z_{1-\alpha / 2} \sqrt{\frac{\wh\bS(\wh\bo)_{jj}}{n}})$.
\end{corollary}

Proof of Corollary \ref{cor:ci} follows from the asymptotic normality of the \PSPA~estimator in Corollary \ref{cor:asy.normal.est.w}. Since we plug in the consistent estimator for each element in $\bS(\bo^{\textnormal{opt}})$, Slutsky's theorem implies that $\wh\bS(\wh\bo^{\textnormal{opt}}) \coP \bS(\bo^{\textnormal{opt}})$, and therefore the condition in Corollary \ref{cor:ci} is also satisfied by our algorithm.

The above two corollaries ensure the validity of the inference for our algorithm. We next demonstrate that our method achieves element-wise variance reduction when compared to all baseline methods.

\begin{proposition}\label{prop:opt}
Suppose $\sqrt{n}(\wh\bt_{\PSPA}(\bo)-\bt) \coD N(\0,\bS(\bo))$,  given $\bo^{\textnormal{opt}}$ defined by equation \ref{eq:omega.opt.explicit},  $\bS(\bo^{\textnormal{opt}})_{jj} \leq \bS(\bo)_{jj}$ for all $j \in \{1,\cdots,q\}.$
\end{proposition}

Proof of Proposition \ref{prop:opt} is contained in Appendix \ref{proof:prop:opt}. By setting the weighting vector $\bo$ to $\bo^{\textnormal{C}} = [0,\dots,0]\trans$ or $\bo^{\PPI} = [1,\dots,1]\trans$, our method reduces to classical method and \PPI, respectively. Therefore, \PSPA~guarantees an element-wise reduction in the asymptotic variance compared to both the classical method and \PPI, for arbitrary ML models.

\subsection{Examples: linear and logistic regression}\label{subsec:ex.regression}
Here, we present two examples to illustrate the intuitions behind the 
\PSPA~estimation procedure. We begin with the estimation of coefficients in a linear regression model, where the resulting estimator is expressed in a closed form.

\begin{example}[Linear Regression]  We consider the ordinary least squares problem, where $\bpsi(y,\x;\bt) = \x(y - \x\trans\bt)$. Therefore, the estimand is $\bt_0 = \E[\X\X\trans]^{-1}\E[\X Y]$. From equation~\eqref{eq:pspa.estimator}, the \PSPA~estimator is the solution to the equation 
\begin{align}
\small
\meann \x_i(y_i - \x_i\trans \bt) 
+
\dfrac{1}{N}\sum_{i=n+1}^{n+N}\diag(\bo) {\x}_i\{\wh f -\x_i\trans\bt\}
- 
\meann \diag(\bo) \x_i\{\wh f - \x_i\trans\bt\} = \0,
\end{align}
yielding a closed-form solution for \PSPA~linear regression
\begin{small}
    \begin{align}
    \wh\bt\prop=& \left\{
    \meann\x_i \x_i\trans + \diag(\wh\bo^{\textnormal{opt}}) \left[\dfrac{1}{N}\sum_{i=n+1}^{n+N} {\x}_i \x_i\trans
     - \meann \x_i\x_i\trans
    \right]
    \right\}\inv
  \\
  &\left\{
    \meann \x_iy_i + \diag(\wh\bo^{\textnormal{opt}}) \left[\dfrac{1}{N}\sum_{i=n+1}^{n+N} \x_i {\wh f}(\z_i) - \meann \x_i\wh f
    \right]
    \right\},
    \end{align}   
\end{small}
where $\wh\bo^{\textnormal{opt}}$ is a consistent estimator of optimal $\bo^{\textnormal{opt}}$, obtainable through Algorithm \ref{alg:pspa:y}. 
\end{example}
The second example is logistic regression, which lacks a closed-form solution but can be solved using Algorithm \ref{alg:pspa:y}.
\begin{example}[Logistic Regression]
For logistic regression, $\bpsi(y,\x;\bt) = -\x y+ \x \gamma_{\bt}(\x)$, where $\gamma_{\bt}(\x) = 1 /\left\{1+\exp \left(-\x\trans \bt\right)\right\}$. Using estimation equation~\eqref{eq:pspa.estimator}, the \PSPA~estimator $\wh\bt\prop $ is the solution to the equation 
\begin{align}
\small
\meann \left\{-\x_iy_i + \x_i\gamma_{\bt}(\x_i)\right\} +
\diag(\bo) \left\{ \dfrac{1}{N}\sum_{i=n+1}^{n+N}  [-{\x}_i \wh f(\z_i) + {\x}_i\gamma_{\bt}(\x_i)]  - \meann [-\x_i\wh f + \x_i\gamma_{\bt}(\x_i)] \right\} = \0
\end{align}
\end{example}

\section{Relationship with Efficient Influence Function}
\label{sec:semi}
To connect our method with the literature on semiparametric efficiency \citep{bickel1993efficient,tsiatis2006semiparametric}, in this section, we state the efficient influence function (EIF) and the semiparametric efficiency bound for the estimation problem~\eqref{eq:esteq} (Proposition \ref{prop:eif}) and connect it with our proposed \PSPA~estimator.  To ease notation, we introduce a random variable $R$ to indicate the labeling; $r_1=\cdots = r_n =1$ and $r_{n+1} = 
\cdots =r_{n+N} = 1$. In this section, suppose $\pi \equiv \pr(R=1) = n/(n+N)$ is a fixed constant between zero and one. Note that the introduction of $R$ is for notational simplicity and we make no inference on it. 

\begin{proposition}[Efficient Influence Function]
\label{prop:eif}
The EIF for estimating $\bt$ is 
\begin{align}
    \bphi(y_i,\x_i,\bt)  = \dfrac{r}{\pi}\A\inv [\bpsi(y,\x;\bt) 
    -\E\{\bpsi(Y,\x;\bt)\mid\x,\z\}] + \A\inv \E\{\bpsi(Y,\x;\bt)\mid\x,\z\}.
\end{align}

\end{proposition}
The proof is contained in Appendix \ref{proof:prop.eif}. Guided by Proposition \ref{prop:eif}, an EIF-based estimator would be the solution to the empirical equation,
\begin{align}\label{eq:eif}
    \0  = \dfrac{1}{n+N}\sum_{i=1}^{n+N}\bphi(y_i,\x_i,\teff).
\end{align}

With the derived EIF, the EIF-based estimator is asymptotically normal and achieves the semiparametric efficiency bound by section 8.4 of \cite{molenberghs2014handbook}.
\begin{corollary}[Semiparametric Efficiency Bound]\label{cor:eif.asy.normal}
     Under Conditions (C1) and (C3), assume \\ $\E\{\sup_{\bt\in\bT}\bphi(Y,\X,\bt)\}<\infty$ and equation~\eqref{eq:eif} has a unique solution, then as $n\to\infty$ and $N\to\infty$,
     \begin{align}
     \sqrt n (\teff - \bt_0) \coD  \mathcal{N}(\0,\E\{\bphi(Y,\X,\bt_0)\bphi(Y,\X,\bt_0)\trans\pi\}).
     \end{align}
\end{corollary}  

If the nuisance function $\E\{\bpsi(Y,\x;\bt)\mid\x,\z\}$ that appears in the EIF can be correctly specified or well approximated, then an EIF-based estimator  will achieve the semiparametric efficiency bound. However, this is nearly impossible in reality with limited knowledge of the data generation process or a small computational budget. Instead, it's typical to construct an estimator with nuisance functions plugged in as their estimates. We refer to such estimators as ``EIF$\star$-based'', because the influence function of the resulting estimator, denoted by $\teffs$, may or may not be the efficient influence function.

In our setting, the nuisance function $\E\{\bpsi(Y,\x;\bt)\mid\x,\z\}$ can be estimated with $\E\{\bpsi(\wh f,\x;\bt)\mid\x,\z\}$. Upon simple calculation, we can write the resulting $\teffs$ as the solution to
\begin{align}\label{eq:effs.ee}
\small
   \meann \A\inv\bpsi(y_i,\x_i;\bt) + (1-\pi)\A\inv  \left\{-\meann\bpsi(\wh f,\x_i;\bt) + \dfrac{1}{N}\sum_{i=n+1}^{n+N} \bpsi(\wh f,\x_i;\bt)\right\}= \0.
\end{align}
Similar to the equation~\eqref{eq:pspa.estimator} that the \PSPA~estimator solves, the EIF$\star$-based estimator~\eqref{eq:effs.ee} solves an equation with two parts: one term that can produce a consistent estimator with labeled data only, and the other term that utilizes predictions $\wh f$, which always have a mean of zero. If one multiples both sides of~\eqref{eq:pspa.estimator} with matrix $\A\inv$, then~\eqref{eq:effs.ee} corresponds to that equation weight equals to a vector of $1-\pi$.
In estimating the outcome mean (Section \ref{subsec:ex.mean}) where $\A$ is essentially the scalar one, the EIF$\star$-based estimator is the \PSPA~estimator with weight being $1-\pi= N/(n+N)$. 
The difference between~\eqref{eq:pspa.estimator} and~\eqref{eq:effs.ee} clarifies one of the main distinctions between the proposed \PSPA~estimator and an EIF-based estimator. The former seeks variance reduction in a data-adaptive way, while the latter attains minimum variance at correct specifications. In real-world situations, the proposed \PSPA~estimator can be much more efficient in that it utilizes information in a data-adaptive fashion. Numeric comparisons are provided in Section \ref{sec:simulation}.

\section{Extensions}\label{sec:extend}
Next, we extend \PSPA~to address more general scenarios. Specifically, Section \ref{subsec:extension.xhat} outlines an inference procedure for fully labeled outcomes with partially labeled covariate $\x$, and Section \ref{subsec:extension.xhat.yhat} discusses cases where both $y$ and $\x$ are partially labeled.

\subsection{ML-predicted covariates}\label{subsec:extension.xhat}
We adjusted the previously described setup by modifying it to use ML to predict the covariate $\X$ instead of the outcome $Y$. Our data can be divided into two parts: $\calL\cup\calU$. The $\calL$ part includes data points $(y_i,\x_i,\z_i)$ for $i=1,\cdots,n$ and the $\calU$ part includes data points $(y_i,\z_i)$ for $i=n+1,\cdots,N+n$. In addition, we use an external prediction algorithm with the notation $\wh q(\cdot)$. This algorithm is applied to $\z$ to generate the predicted value $\wh q(\z)$ of the covariate $\X$. However, it is important to note that $\wh q$ may produce inaccurate or biased predictions of $\X$. The estimand is also defined by the estimating equation~\eqref{eq:esteq}. Similar to our proposal to handle the ML-predicted outcome, we propose our \PSPA~estimator $\wh\bt_{{\PSPA^{'}}}(\bo)$ with ML-predicted covariates that solves the equation
\begin{align}\label{eq:pspa.estimator.x}
\Psi_{\PSPA*}^\bo(\bt) :=\meann\bpsi(y_i,\x_i;\bt) + \diag(\bo) \left\{-\meann\bpsi(y_i,\wh q;\bt) + \dfrac{1}{N}\sum_{i=n+1}^{n+N} \bpsi(y_i,\wh q;\bt)\right\} = \0,
\end{align}

Denote $\A = \E[\partial\bpsi(Y,\X;\bt)/\partial\bt]$, $\M_1 = \myVar[\bpsi(Y,\X;\bt)]$,
$\M_2' = \myVar[\bpsi( Y,\wh q;\bt_0)]$, 
$\M_3' = \myVar[\bpsi(Y,\wh q;\bt_0)]$, 
$\M_4' = \myCov[\bpsi(Y,\X;\bt_0),\bpsi(Y,\wh q;\bt_0)]$. Upon calculations similar to Theorem~\ref{thm:asy.normal.pspa}, the $j$-th diagonal element of the asymptotic variance of $\wh\bt_{{\PSPA^{'}}}(\bo)$ is
\begin{align}
    \bS'_{jj}(\bo) 
    &=
    \omega_j^2[\A\inv(\M_2'+\rho\M_3')\A\inv]_{jj} - 2\omega_j[\A\inv\M_4'\A\inv]_{jj} + [\A\inv\M_1'\A\inv]_{jj},
\end{align}
which achieve its minimum with $\omega_j^{\textnormal{opt}} = \frac{[\A\inv\M_4'\A\inv]_{jj}}{[\A\inv(\M_2'+\rho\M_3')\A\inv]_{jj}}$. We modify our Algorithm \ref{alg:pspa:y} to the following algorithm to incorporate ML-predicted covariates.

\begin{algorithm}[H]
  \caption{\PSPA~estimation with ML-predicted covariates}
  \label{alg:pspa:x}
  \begin{algorithmic}[1]
  \Require Data $\calL\cup\calU$, pre-trained ML model $\wh f$, error rate $\alpha \in (0, 1)$.
  \State Obtain the classical estimator $\wh\bt_{\text{C}}$ by solving $\meann\bpsi(y_i,\x_i;\bt) = \0$.
  \State Obtain the optimal weighting vector $\wh\bo^{\textnormal{opt}} = [\wh\omega_1^{\textnormal{opt}},\dots,\wh\omega_q^{\textnormal{opt}}]\trans$ by
$$
\wh\omega_j^{\textnormal{opt}} =
\begin{cases}
\min(\frac{[\wh\A_{\text{C}}\inv\wh\M_{4,{\text{C}}}^{'}\wh\A_{\text{C}}\inv]_{jj}}{[\wh\A_{\text{C}}\inv(\wh\M_{2,{\text{C}}}^{'}+\rho\wh\M_{3,{\text{C}}}^{'})\wh\A_{\text{C}}\inv]_{jj}}, \frac{\lambda_{{\min, +}}[\frac{1}{n}\nabla\bpsi(y_i,\x_i;\bt)]}{\lambda_{\max}[\frac{1}{n}\nabla\bpsi(y_i,\wh q;\bt)]}) &\quad\text{if } \frac{[\wh\A_{\text{C}}\inv\wh\M_{4,{\text{C}}}^{'}\wh\A\inv]_{jj}}{[\wh\A_{\text{C}}\inv(\wh\M_{2,{\text{C}}}^{'}+\rho\wh\M_{3,{\text{C}}}^{'})\wh\A_{\text{C}}\inv]_{jj}} >0  \\
\max(\frac{[\wh\A_{\text{C}}\inv\wh\M_{4,{\text{C}}}^{'}\wh\A_{\text{C}}\inv]_{jj}}{[\wh\A_{\text{C}}\inv(\wh\M_{2,{\text{C}}}^{'}+\rho\wh\M_{3,{\text{C}}}^{'})\wh\A_{\text{C}}\inv]_{jj}}, 0) &\quad\text{if } \frac{[\wh\A_{\text{C}}\inv\wh\M_{4,{\text{C}}}^{'}\wh\A\inv]_{jj}}{[\wh\A_{\text{C}}\inv(\wh\M_{2,{\text{C}}}^{'}+\rho\wh\M_{3,{\text{C}}}^{'})\wh\A_{\text{C}}\inv]_{jj}} \leq 0  \\
\end{cases}
$$

for all  $j \in 1, \dots, q$. Here, $\wh\A_{\text{C}}$, $\wh\M_{1,{\text{C}}}^{'}$, $\wh\M_{2,{\text{C}}}^{'}$, $\wh\M_{3,{\text{C}}}^{'}$, $\wh\M_{4,{\text{C}}}^{'}$ are sample analogs of $\A$, $\M_1^{'}$, $\M_2^{'}$, $\M_3^{'}$, $\M_4^{'}$ for $\Psi_{\PSPA*}^\bo(\bt)$ with $\wh\bt_{\text{C}}$ plugged in. $\lambda_{{\min, +}}[\M]$ and $\lambda_{\max}[\M]$ represent the smallest non-negative and largest eigenvalue of matrix $\M$, respectively.
\State Obtain the $\PSPA^*$~estimator with a one-step update:
$$
   \wh\bt_{{\PSPA^{'}}} = \wh\bt_{\text{C}} - \left[\nabla\Psi_{\PSPA^{'}}^{\wh\bo^{\textnormal{opt}}}(\wh\bt_{\text{C}})\right]\inv
   \Psi_{\PSPA^{'}}^{\wh\bo^{\textnormal{opt}}}(\wh\bt_{\text{C}}), \text{ where } \nabla\Psi_{\PSPA^{'}}^{\wh\bo^{\textnormal{opt}}} = \partial\Psi_{\PSPA^{'}}^{\wh\bo^{\textnormal{opt}}}(\bt)/\partial\bt\vert_{\bt = \wh\bt_{\text{C}}}.
$$

\State Obtain the asymptotic variance of $\wh\bt_{{\PSPA^{'}}}$:
$$
    \wh\bS'(\wh\bo^{\textnormal{opt}}) = \wh\A_{\PSPA^{'}}\inv[\wh\M_{1,\PSPA}^{'} + \diag(\wh\bo^{\textnormal{opt}})(\wh\M_{2,\PSPA}^{'} + \rho\wh\M_{3,\PSPA}^{'})\diag(\bo) - 2\diag(\wh\bo^{\textnormal{opt}})\wh\M_{4,\PSPA}^{'}]\wh\A_{\PSPA^{'}}\inv,
$$
where $\wh\A_\PSPA^{'}$, $\wh\M_{1,\PSPA}^{'}$, $\wh\M_{2,\PSPA}^{'}$, $\wh\M_{3,\PSPA}^{'}$, $\wh\M_{4,\PSPA}^{'}$ are sample analogs of $\A$, $\M_1^{'}$, $\M_2^{'}$, $\M_3^{'}$, $\M_4^{'}$ for $\Psi_{\PSPA^{'}}^\bo(\bt)$ with $\wh\bt_\PSPA$ plugged in.

  \Ensure $\PSPA^{'}$~estimator $\wh\bt_{{\PSPA^{'}}}$, standard error  $\sqrt{\frac{\wh\bS'(\wh\bo^{\textnormal{opt}})_{jj}}{n}}$, $\alpha$-level confidence interval $\mathcal{C}_{\alpha,j}^{\PSPA^{'}}=(\wh\bt_{\PSPA^{'}_j} \pm z_{1-\alpha / 2} \sqrt{\frac{\wh\bS'(\wh\bo^{\textnormal{opt}})_{jj}}{n}})$, and (two-sided) p-value $ 2 ( 1 - \Phi(| \frac{\wh{\bt}_{\PSPA^{'}_j}}{\sqrt{\frac{\wh\bS'(\wh\bo^{\textnormal{opt}})_{jj}}{n}})} | ) )$ for the $j$-th coordinate. Here, $\Phi$ is the CDF of the standard normal distribution.
   \end{algorithmic}
\end{algorithm}

Corollary \ref{cor:x} states the asymptotic distribution that enables interval estimation with its proof follows from the proof of Theorem \ref{thm:asy.normal.pspa}, Corollary \ref{cor:asy.normal.est.w} and \ref{cor:ci}, and thus omitted.

\begin{corollary}\label{cor:x}
    Under Conditions (C1), (C2) on equation \ref{eq:pspa.estimator.x}, and (C3)-(C4), assuming $\frac{n}{N} \rightarrow \rho$ as $n\to\infty$ and $N \rightarrow \infty$, and $\wh \bo \coP \bo$, then the proposed estimator $\tprop(\wh \bo)$ converges to $\bt_0$ in probability. Assuming additionally Condition (C5), we have that as $n\to\infty$ and $N\to\infty$,
\begin{align}\label{eq:asy.normal}
    \sqrt n(\wh\bt'\prop({\wh \bo}) - \bt_0)\coD \mathcal{N}(\0,\bS'(\bo)).
\end{align}
where $\bS'(\bo) = \A\inv\V'(\bo)\A\inv$, $\A = \E[\partial\bpsi(Y,\X;\bt)/\partial\bt]$,
$\V'(\bo) = \M_1 + \diag(\bo)(\M_2' + \rho\M_3')\diag(\bo) - 2\diag(\bo)\M_4'$, $\M_1 = \myVar[\bpsi(Y,\X;\bt)]$,
$\M_2' = \myVar[\bpsi( Y,\wh q;\bt_0)]$, 
$\M_3' = \myVar[\bpsi(Y,\wh q;\bt_0)]$, \\
$\M_4' = \myCov[\bpsi(Y,\X;\bt_0),\bpsi(Y,\wh q;\bt_0)]$. With $\wh\bS'(\wh\bo) \coP \bS'(\bo)$, 
\begin{align}
    \lim_{n \rightarrow \infty}\mathbb{P}(\theta_{j}\in\mathcal{C}_{\alpha,j}^{{'}\wh\bo})\geq 1-\alpha,\ j=1,\cdots,q,
\end{align}
 where $\mathcal{C}_{\alpha,j}^{{'}\wh\bo}=(\wh\bt_{\textnormal{\PSPA}_j}'(\wh\bo) \pm z_{1-\alpha / 2} \sqrt{\frac{\wh\bS'(\wh\bo)_{jj}}{n}})$, $1-\alpha\in(0,1)$ is the confidence level, and $\theta_{j}$ is the $j$-th coordinate of the parameter.
\end{corollary}

\begin{remark}
Condition (C2) requires equation \ref{eq:pspa.estimator.x} to have a unique solution. To ensure this, we impose a constraint on $\bo$. Equation \ref{eq:pspa.estimator.x} has a unique solution if its derivative, $ \meann[\nabla\bpsi(y_i,\x_i;\bt) - \diag(\bo)\nabla\bpsi(y_i,\wh q;\bt)] + \dfrac{1}{N}\sum_{i=n+1}^{n+N} \nabla\bpsi(y_i,\wh q;\bt), $
is positive semi-definite. The second term, $\dfrac{1}{N}\sum_{i=n+1}^{n+N} \nabla\bpsi(y_i,\wh q;\bt)$, is already positive semi-definite, so we only need to ensure the first term is positive semi-definite.
According to Lemma \ref{lem:eigen} in the appendix, a sufficient condition for this is:
$0 \leq \omega_j \leq \dfrac{\lambda_{{\min, +}}[\meann\nabla\bpsi(y_i,\x_i;\bt)]}{\lambda_{\max}[\meann\nabla\bpsi(y_i,\wh q;\bt)]}$
for all $j = 1, \cdots, q$, where $\lambda_{{\min, +}}[\M]$ and $\lambda_{\max}[\M]$ represent the smallest positive and largest eigenvalue of matrix $\M$, respectively. By imposing this constraint on the weighting vector in the algorithm, we ensure that condition (C2) holds. In practice, one can relax this constraint by substituting the prefixed $\bo$ into the derivative and confirming its positive semi-definiteness. If necessary, one can iteratively reduce the value of each positive $\omega_j$ (or increase negative $\omega_j$)w until the derivative achieves positive semi-definiteness.
\end{remark}

Next, we state the theoretical guarantees of \PSPA~for element-wise variance reduction compared with the classical approach with weighting vector $\bo^{\textnormal{C}} = [0,\dots,0]\trans$.

\begin{corollary}\label{prop:opt:x}
Suppose $\sqrt{n}(\wh\bt_{\PSPA'}(\bo^{\textnormal{C}})-\bt) \coD N(\0, \bS(\0))$. Denote $\hat\bo^{\textnormal{opt}} = [\wh\omega_1^{\textnormal{opt}},\dots,\wh\omega_q^{\textnormal{opt}}]\trans$, where 
\begin{align}
\wh\omega_j^{\textnormal{opt}} =
\begin{cases}
\min(\frac{[\wh\A\inv\wh\M_{4,{\text{C}}}^{'}\wh\A\inv]_{jj}}{[\wh\A\inv(\wh\M_{2,{\text{C}}}^{'}+\rho\wh\M_{3,{\text{C}}}^{'})\wh\A\inv]_{jj}}, \frac{\lambda_{{\min, +}}[\frac{1}{n}\nabla\bpsi(y_i,\x_i;\bt)]}{\lambda_{\max}[\frac{1}{n}\nabla\bpsi(y_i,\wh q;\bt)]}) &\quad\text{if } \frac{[\wh\A\inv\wh\M_{4,{\text{C}}}^{'}\wh\A\inv]_{jj}}{[\wh\A\inv(\wh\M_{2,{\text{C}}}^{'}+\rho\wh\M_{3,{\text{C}}}^{'})\wh\A\inv]_{jj}} >0  \\
\max(\frac{[\wh\A\inv\wh\M_{4,{\text{C}}}^{'}\wh\A\inv]_{jj}}{[\wh\A\inv(\wh\M_{2,{\text{C}}}^{'}+\rho\wh\M_{3,{\text{C}}}^{'})\wh\A\inv]_{jj}}, 0) &\quad\text{if } \frac{[\wh\A\inv\wh\M_{4,{\text{C}}}^{'}\wh\A\inv]_{jj}}{[\wh\A\inv(\wh\M_{2,{\text{C}}}^{'}+\rho\wh\M_{3,{\text{C}}}^{'})\wh\A\inv]_{jj}} \leq 0  \\
\end{cases}
\end{align}
for all $j \in 1, \dots, q$. Suppose $\hat\bo^{\textnormal{opt}} \coP \bo^{\textnormal{opt}}$, then $\bS(\bo^{\textnormal{opt}})_{jj} \leq \bS(\bo^{\textnormal{C}})_{jj}$.
\end{corollary}

The proof is contained in Appendix \ref{proof:prop:opt:x}.

\subsection{ML-predicted outcome and covariates}\label{subsec:extension.xhat.yhat}
Next, we explore a situation where the outcome and covariates are not directly observed in the unlabeled data. Instead, we employ machine learning techniques to predict both. Additionally, we possess much smaller data with both outcome and covariates measured. This full data can be categorized into two segments: $\calL\cup\calU$. The first segment, $\calL$, includes the data points $(y_i,\x_i,\z_i)$ for $i=1,\cdots,n$, while the second segment, $\calU$, comprises the instances of $\z_i$ for $i=n+1,\cdots,N+n$, and $N >> n$.
The variable $\z$ is to predict both the outcome and the covariates. The ML-predicted outcome and covariates are represented by $\wh f$ and $\wh q(\z)$, respectively, where $\wh f$ and $\wh q$ are independently obtained.
The \PSPA~estimator for this situation is to solve the equation,
\begin{align}\label{eq:pspa.estimator.xy}
\meann\bpsi(y_i,\x_i;\bt) + \diag(\bo) \left\{-\meann\bpsi(\wh f,\wh q;\bt) + \dfrac{1}{N}\sum_{i=n+1}^{n+N} \bpsi(\wh f,\wh q;\bt)\right\} = \0,
\end{align}
Replacing the estimating equation in Algorithm \ref{alg:pspa:x} with Equation \ref{eq:pspa.estimator.xy} gives an algorithm for estimation and statistical inference for this task. The consistency and asymptotic normality of the resulting estimator can be established by the proof of Theorem \ref{thm:asy.normal.pspa}.

\section{Simulations}\label{sec:simulation}
We conduct simulations to evaluate the performance of \PSPA~using linear and logistic regression across two settings: one with ML-predicted labels and another with ML-predicted covariates. In the ML-predicted labels setting, we considered classical, \PPI, and EIF*-based methods. The EIF*-based method employs a weighting vector defined as $\bo^{\textnormal{EIF*}} = \left[\frac{N}{N+n},\dots,\frac{N}{N+n}\right]\trans$, where $N$ and $n$ represent the sample sizes of unlabeled and labeled data, respectively. For the ML-predicted covariates setting, we used classical and imputation-based methods as baselines. We present the implementation details in Section \ref{subsec:simulation.setup} and the simulation results in Section \ref{subsec:simulation.result}.

\subsection{Implementation details}\label{subsec:simulation.setup}
In all simulations, the ground truth coefficients are obtained using \(5 \times 10^4\) samples. The labeled data is with 500 samples and the unlabeled data is with 500, 1500, 2500, 5000, or 10000 samples for different settings. A pre-trained random forest with 100 trees to grow is obtained from a hold-out data with 1000 samples. All simulations are repeated 1000 times.

\subsubsection{ML-predicted labels}
We simulate the labels $Y_i$ and covariates $X_{1i},\dots,X_{50i}$ for linear regression setting by
$
  X_{1i},\dots,X_{50i} \overset{\text{i.i.d}}{\sim} \mathcal{N}(0, 1), Z_i \sim \mathcal{N}(0, 1),
  \theta_1,\dots,\theta_{10} = \frac{0.1}{\sqrt{10}};,\theta_{11},\dots,\theta_{50} = 0,
  Y_i = \sum_{k=1}^{50}X_{ki}\theta_k + rZ_i + \epsilon_i, \epsilon_i \sim \mathcal{N}(0, \tau_\epsilon) \text{ where } \tau_\epsilon$ such that $\operatorname{Var}(Y_i) = 1,
$
where $r$ is set to be 0.8 for settings with different sample size of unlabeled data and 0, 0.2, 0.4, 0.6, or 0.8 for settings with different imputation accuracy. We used the $X_{1i}, \dots, X_{50i}$, and $Z_i$ as inputs to train the random forest model, which aimed to predict $Y_i$ in unlabeled data. For logistic regression, we used the same data-generating process except that we generate the label $\tilde{Y}_i$ by $\tilde{Y}_i = \mathds{1}(Y_i > \text { median }(Y_i))$. Our parameter of interest is the regression coefficient for $X_{1i}$.

\subsubsection{ML-predicted covariates}
We simulate the labels $Y_i$ and covariates $X_{1i},\dots,X_{10i}$ for linear regression setting by
$
  X_{1i},\dots,X_{10i} \overset{\text{i.i.d}}{\sim} \mathcal{N}(0, 1), Z_i \sim \mathcal{N}(0, 1),
  \theta_1 = 0.1, \theta_{2},\dots,\theta_{10} = 0,
  Y_i = \sum_{k=1}^{10}X_{ki}\theta_k + \epsilon_i, \epsilon_i \sim \mathcal{N}(0, \tau_\epsilon) \text{ where } \tau_\epsilon$ is set to be the value such that $\operatorname{Var}(Y_i) = 1, Z_i = 0.1Y_i + rX_{1i} + \delta_i \text{ where } \delta_i \sim \mathcal{N}(0, \tau_\delta)$ where $\tau_\delta$ is set to be the value such that $\operatorname{Var}(Z_i) = 1$
where $r$ is set to be 0.8 for settings with different sample size of unlabeled data and 0, 0.2, 0.4, 0.6, or 0.8 for settings with different imputation accuracy. We employed the variables $Z_i$ as inputs to train the random forest model, which aimed to predict $X_{1i}$ in unlabeled data. For logistic regression, we used the same data-generating process except that we generate the label $\tilde{Y}_i$ by $\tilde{Y}_i = \mathds{1}(Y_i > \text { median }(Y_i))$. We are interested in the regression coefficient for $X_{1i}$. 
\begin{figure}[H]
    \centering
    \includegraphics[width = 1\linewidth]{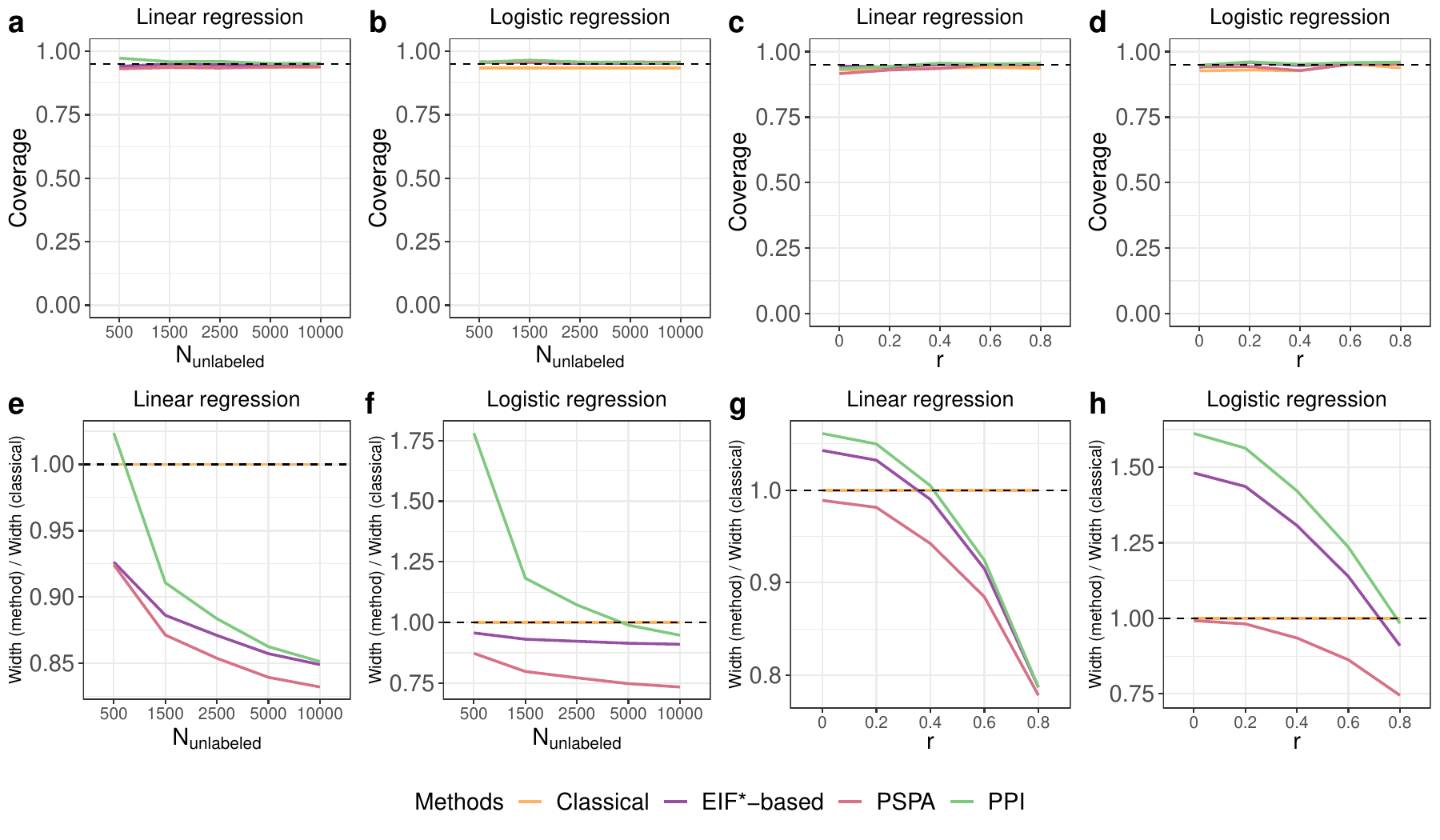}
    \caption{\textbf{Coverage of the confidence interval and relative ratio of its width compared to the classical method for linear and logistic regression.} ML is used to predict the labels. Panels (a)-(d) show the coverage of the confidence interval. Panels (e)-(h) show the relative ratio of the width of the confidence interval in comparison with the classical method. Panels (a), (b), (e), and (f) correspond to settings with varying sample sizes of unlabeled data. Panels (c), (d), (g), and (h) correspond to settings with different levels of imputation accuracy. The dashed line represent $y = 0.95$ in (a)-(d) and $y=1$ in (e)-(h).}
    \label{fig:sim_Y_PPI}
\end{figure}

\subsection{Results} \label{subsec:simulation.result}
\subsubsection{ML-predicted labels}
Figure \ref{fig:sim_Y_PPI} shows the simulation results for the setting of ML-predicted labels. All evaluated methods achieved confidence interval coverage rates close to 95\%, indicating inference validity across different unlabeled sample sizes and different levels of ML prediction accuracy. \PPI~and EIF*-based methods may be less efficient than classical methods when the ML has poor prediction accuracy or limited unlabeled data. \PSPA~has narrower confidence intervals compared to classical method and other baseline methods, especially as the unlabeled sample size and ML prediction performance increase.

\subsubsection{ML-predicted covariates}
\begin{figure}[H]
    \centering
    \includegraphics[width = 1\linewidth]{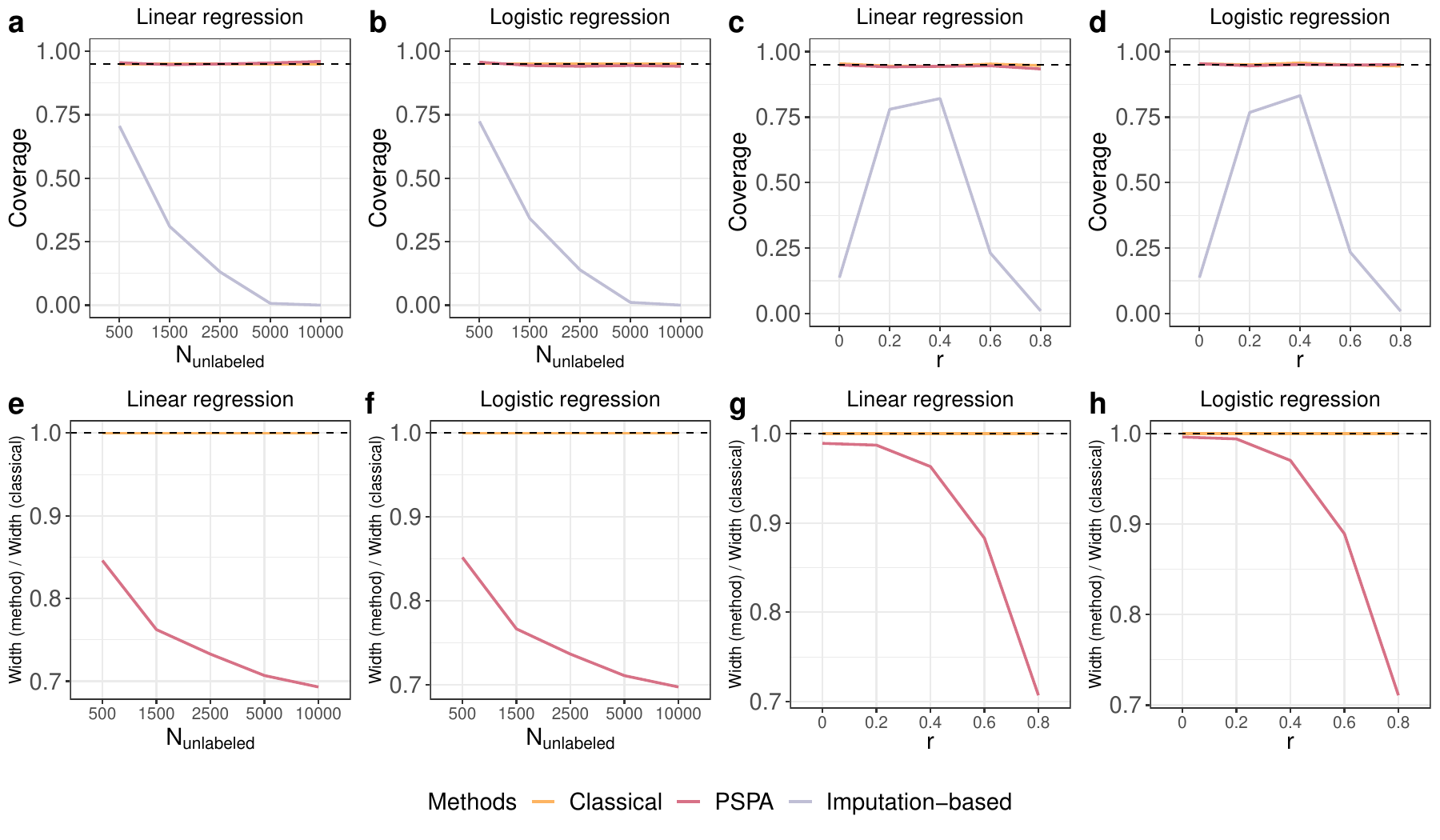}
        \caption{\textbf{Coverage of the confidence interval and relative ratio of its width compared to the classical method for linear and logistic regression.} ML is used to predict the covariates. Panels (a)-(d) show the coverage of the confidence interval. Panels (e)-(h) show the relative ratio of the width of the confidence interval in comparison with the classical method. Panels (a), (b), (e), and (f) correspond to settings with varying sample sizes of unlabeled data. Panels (c), (d), (g), and (h) correspond to settings with different levels of imputation accuracy. The dashed line represent $y = 0.95$ in (a)-(d) and $y=1$ in (e)-(h).}
    \label{fig:sim_X}
\end{figure}

Figure \ref{fig:sim_X} shows the simulation results for the setting with ML-predicted covariates. The imputation-based method fails to achieve the correct confidence interval coverage. Both classical and \PSPA~methods lead to coverage rates around 95\%, which suggests  valid inference results. \PSPA~has narrower confidence intervals compared to the classical method, regardless of the unlabeled sample size and the accuracy of the ML model.

\section{Data Applications}\label{sec:real}
\subsection{Sex-differentiated gene expression}
We used the \PSPA~method to assess the effect of biological sex on gene expression across 44 human tissues using GTEx data. We aim to use an ML algorithm called hypergraph factorization (HYFA) to impute the missing gene expression \citep{vinas2023hypergraph} in the uncollected tissue. The imputed gene expression is then used in \PSPA~to increase the statistical power to identify sex-biased genes. We compared \PSPA~with classical, \texttt{PPI}, and EIF$\star$-based approaches.  In particular, we processed the GTEx-v8 data following the GTEx pipeline \citep{gtex2015genotype}. We then used HYFA to predict transcript levels for each uncollected tissue for each individual in the data, resulting in a gene expression dataset of $834$ individuals across $44$ tissues. HYFA is a parameter-efficient graph representation learning approach for multi-tissue prediction of gene expression. We used cross-validation to predict expression levels in the labeled data to avoid over-fitting. Prediction accuracy is measured by the correlation between measured and imputed gene expression in the labeled data. We then used linear regression to assess the effect of sex on gene expression while controlling for technical factors including surrogate variables, age, RNA integrity number, and total ischemic time for a sample.

\begin{figure}[H]
    \centering
    \includegraphics[width = 0.7\linewidth]{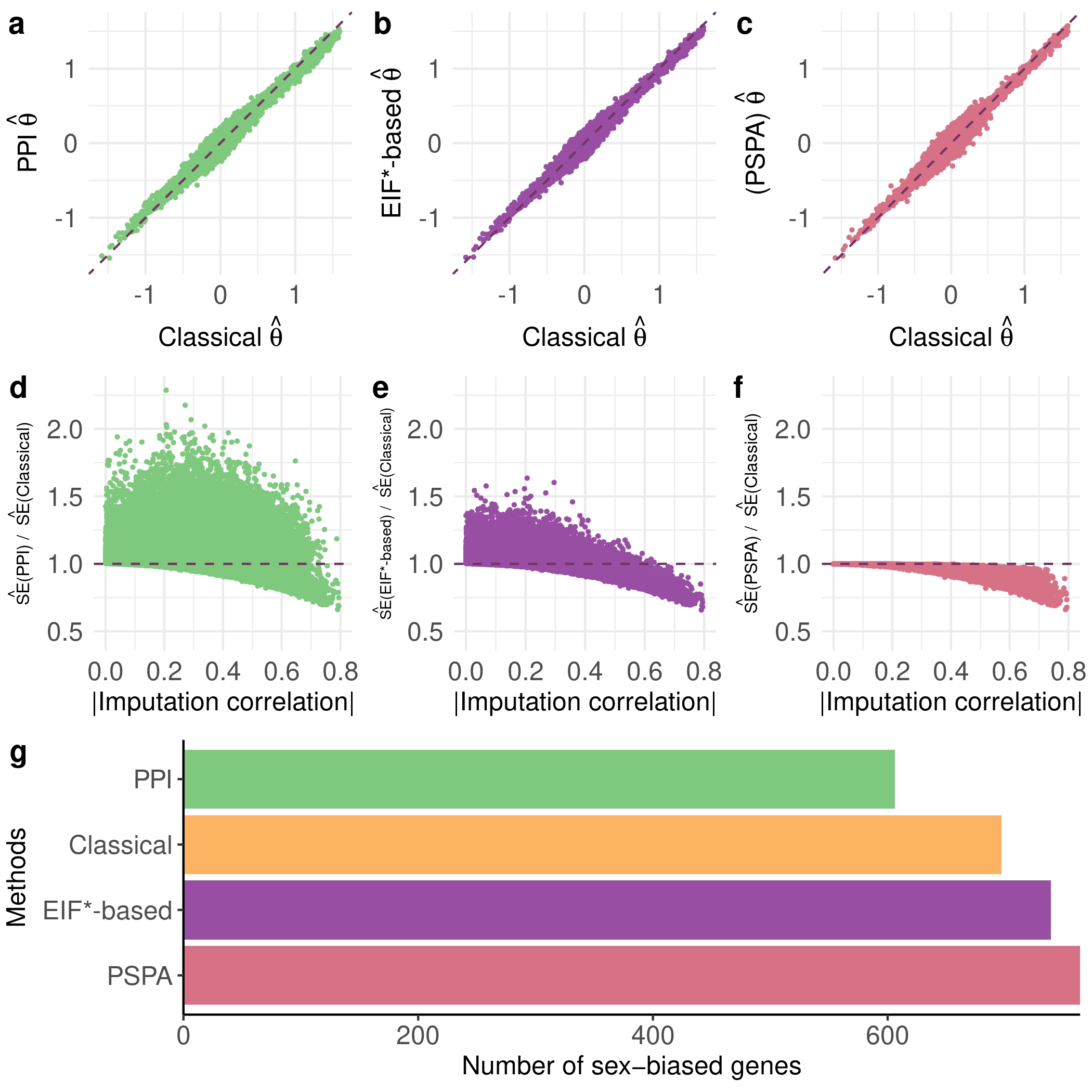}
    \caption{\textbf{Comparison of \PSPA, classical, \texttt{PPI}, and EIF$\star$-based approaches in identifying sex-biased gene expressions using GTEx data.} Each panel illustrates a different aspect of comparison on the y- and x- axes: point estimates between the (a) classical and \texttt{PPI} approaches; (b) classical and EIF$\star$-based approaches; (c) classical and \PSPA~approaches; estimated standard errors between the (d) classical and \texttt{PPI} approaches; (e) classical and EIF$\star$-based approaches; (f) classical and \PSPA~approaches; (g) number of sex-biased genes identified by each of the four approaches. The dashed lines represent $y = x$ in (a)-(c) and $y=1$ in (d)-(f).} 
    \label{fig:gtex}
\end{figure}

Results are shown in Figure \ref{fig:gtex}. Each dot represents the result of the inference for one gene in one tissue in Figure \ref{fig:gtex} (a)-(f). The point estimates of all methods are close to the classical approach (Figure \ref{fig:gtex} (a)-(c)). However, both \texttt{PPI} and EIF$\star$-based are less efficient than the classical approach when the prediction accuracy is low (Figure \ref{fig:gtex} (d)-(e)).  In contrast, the \PSPA~estimator is always no less efficient than the classical approach, regardless of the prediction accuracy (Figure \ref{fig:gtex} (f)). Moreover, \PSPA~identifies more sex-biased genes than other approaches, demonstrating its improved efficiency over alternatives (Figure \ref{fig:gtex} (g)).

\subsection{Risk factors for bone mineral density}
Next, we applied \PSPA~to a multi-dimensional linear regression task. The goal is to identify the associations between dual-energy x-ray absorptiometry (DXA)-derived total bone mineral density (DXA-BMD) and several covariates, including biological sex, age, physical activities (PA), sedentary behavior (SB), smoking status (current smoker or not), and frequency of alcohol intake. DXA-BMD serves as the primary diagnostic marker for osteoporosis and fracture risk in clinical settings. For PA, we assigned individuals three levels (low, medium, and high) score according to the International Physical Activity Questionnaire guidelines. SB was quantified as an integer value representing the combined hours spent driving, using a computer, and watching television.

\begin{table}[H]
    \centering
    \renewcommand{\arraystretch}{1.25}
    \setlength{\tabcolsep}{3pt}
    \caption{\textbf{Comparison of different methods in identifying risk factors for bone mineral density.} Estimate and $\wh{\text{SE}}$ represent the estimated linear regression coefficient and its corresponding standard error for each covariates. $\wh{\text{SE}}$ ratio indicates the ratio of the standard error for a specific method compared with the classical method. PA denotes physical activity, and SB refers to sedentary behavior. The bold font represent the method that gives the smallest $\wh{\text{SE}}$ ratio for each covariate.}
    \label{table:BMD}
    \begin{tabular}{c|c|c|c|c|c|c|c}
        & & Biological sex & Age & PA & SB & Smoking & Alcohol \\ \hline
        \multirow{3}{*}{Classical} & Estimate & -0.616 & -0.190 & 0.019 & 0.040 & 0.006 & -0.008 \\ \cline{2-8}
        & $\wh{\text{SE}}$ & 4.20E-03 & 4.12E-03 & 4.07E-03 & 4.18E-03 & 4.13E-03 & 4.20E-03 \\ \cline{2-8}
        & $\wh{\text{SE}}$ ratio & 1 & 1 & 1 & 1 & 1 & 1 \\ \hline
        \multirow{3}{*}{\PPI} & Estimate & -0.604 & -0.201 & 0.009 & 0.043 & -0.026 & -0.002 \\ \cline{2-8}
        & $\wh{\text{SE}}$ & 5.73E-03 & 6.26E-03 & 6.35E-03 & 6.48E-03 & 6.39E-03 & 6.47E-03 \\ \cline{2-8}
        & $\wh{\text{SE}}$ ratio & 1.363 & 1.517 & 1.560 & 1.553 & 1.549 & 1.541 \\ \hline
        \multirow{3}{*}{EIF*-based} & Estimate & -0.605 & -0.200 & 0.010 & 0.042 & -0.023 & -0.003 \\ \cline{2-8}
        & $\wh{\text{SE}}$ & 5.28E-03 & 5.74E-03 & 5.83E-03 & 5.95E-03 & 5.87E-03 & 5.93E-03 \\ \cline{2-8}
        & $\wh{\text{SE}}$ ratio & 1.257 & 1.391 & 1.433 & 1.424 & 1.423 & 1.412 \\ \hline
        \multirow{3}{*}{\PSPA} & Estimate & -0.614 & -0.181 & 0.010 & 0.040 & -0.008 & -5.15E-07 \\ \cline{2-8}
        & $\wh{\text{SE}}$ & 3.74E-03 & 3.74E-03 & 3.74E-03 & 3.82E-03 & 3.80E-03 & 3.81E-03 \\ \cline{2-8}
        & $\wh{\text{SE}}$ ratio & \textbf{0.892} & \textbf{0.907} & \textbf{0.919} & \textbf{0.914} & \textbf{0.920} & \textbf{0.908} \\ 
    \end{tabular}
\end{table}

We regressed DXA-BMD on these variables using data from the UK Biobank (UKB). In the UKB, DXA-BMD measurements are available for only 10\% of the participants. Therefore, we employed the Softimpute algorithm to impute DXA-BMD values for the remaining 90\% individuals in the unlabeled dataset. To prevent overfitting, cross-validation was applied for the imputation in the labeled data. We consider classical, \PPI, and EIF*-based method as baseline method.

Table \ref{table:BMD} presents the inference results. Both the \PPI~and EIF*-based methods exhibit larger standard errors across all covariates compared to the classical method. In contrast, \PSPA~shows smaller standard errors for all coordinates when compared to the classical method, demonstrating its feature for element-wise variance reduction.

In conclusion, the real data results are consistent with our theoretical and simulation results, demonstrating that our method \PSPA~provides more efficient inference results in real-world post-prediction inference applications.

\section{Comparison with \PPIpp}\label{sec:ppi++}
Finally, we present a comparison between \PPIpp~ \cite{angelopoulos2023ppi++} and our method \PSPA, both theoretically and empirically. First, we note that \PPIpp~can be reviewed as a special case of \PSPA~when each element of the weighting vector $\bo$ is constrained to the same values. \PSPA~ degenerates to \PPIpp~ for one-dimensional estimation task such as mean estimation. Moreover, \PPIpp~cannot be applied to the case where the covariates are predicted by ML instead of the labels. Theoretically, since \PPIpp~can be reviewed as a special case of \PSPA, \PPIpp~is less efficient than \PSPA~by Proposition \ref{prop:opt}. Moreover, \PPIpp~fails to guarantee element-wise variance reduction since it only incorporates one scalar for variance reduction.

Empirically, we compared \PPIpp~with \PSPA~for all experiments detailed in this paper. We present the simulation results in Figure \ref{fig:sim_Y_PPI++} and real data applications in Figure \ref{fig:GTex_PPI++} and Table \ref{table:BMD_PPI++}, respectively. Simulation results show that both \PPIpp~and \PSPA
~have the correct confidence interval coverage. However, \PSPA~has narrower confidence intervals compared with \PPIpp. In real data applications, \PSPA~identifies more sex-biased gene expressions than \PPIpp~in the GTEx example and has a smaller estimated standard error compared to \PPIpp~ for every covariates in the DXA-BMD example. These results are consistent with Proposition \ref{prop:opt}, showing \PSPA~is more efficient than \PPIpp~in post-prediction inference applications.


\section{Discussion}\label{sec:discussion}
We have provided a simple yet powerful method, \PSPA, to improve the efficiency of statistical inference with arbitrary ML predictions. To enable inference, we establish the consistency and asymptotic normality of the proposed estimator and prove its superiority over baseline methods. Through extensive simulations and real data applications, we demonstrate the superiority of the \PSPA~estimator over alternatives. 

This work fits into a variety of scientific problems and unlocks several directions for future research. One can apply the same principle to more complicated data structures and missing mechanisms, for example in cases where a multi-dimensional outcome or covariate is subject to missing. It would be an interesting future work to carefully discuss relationships between $\X$ and $\Z$ and how they may affect the magnitude of the efficiency improvement.

The R codes to implement \PSPA, benchmark methods, and replicate the simulation and real data analysis, is available at \href{https://github.com/qlu-lab/pspa}{\texttt{https://github.com/qlu-lab/pspa}}.


\subsubsection*{Acknowledgments}
We gratefully acknowledge research support from the University of Wisconsin-Madison Office of the Chancellor and the Vice Chancellor for Research and Graduate Education with funding from the Wisconsin Alumni Research Foundation (WARF). We thank the participants of the UW-Madison Statistics Graduate Student Association Student Seminar and the Lu Group Lab Meeting for many useful comments and questions.


\newpage

\appendix
\section{Supplementary Figures and Tables}
\begin{figure}[H]
    \centering
    \includegraphics[width = 1\linewidth]{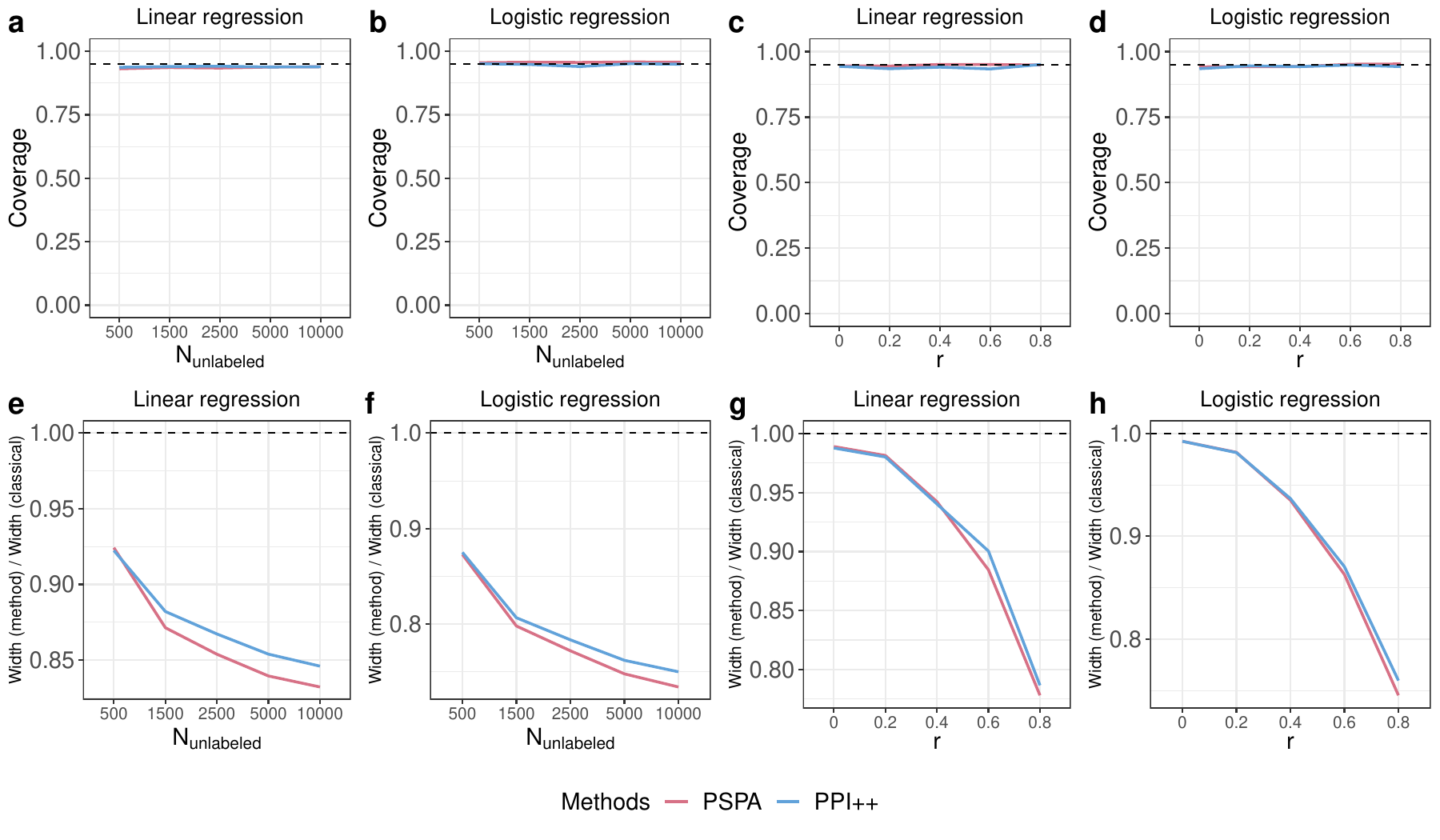}
    \caption{\textbf{Comparison of \PPIpp~and \PSPA~in simulation.} This figure shows the coverage of the confidence interval and the relative ratio of the width of the confidence interval compared to the classical method for linear and logistic regression. ML is used to predict the labels. Panels (a)-(d) show the coverage of the confidence interval. Panels (e)-(h) show the relative ratio of the width of the confidence interval in comparison with the classical method. Panels (a), (b), (e), and (f) correspond to settings with varying sample sizes of unlabeled data. Panels (c), (d), (g), and (h) correspond to settings with different levels of imputation accuracy. The dashed line represent $y = 0.95$ in (a)-(d) and $y=1$ in (e)-(h).}
    \label{fig:sim_Y_PPI++}
\end{figure}

\begin{figure}[H]
    \centering
    \includegraphics[width = 0.7\linewidth]{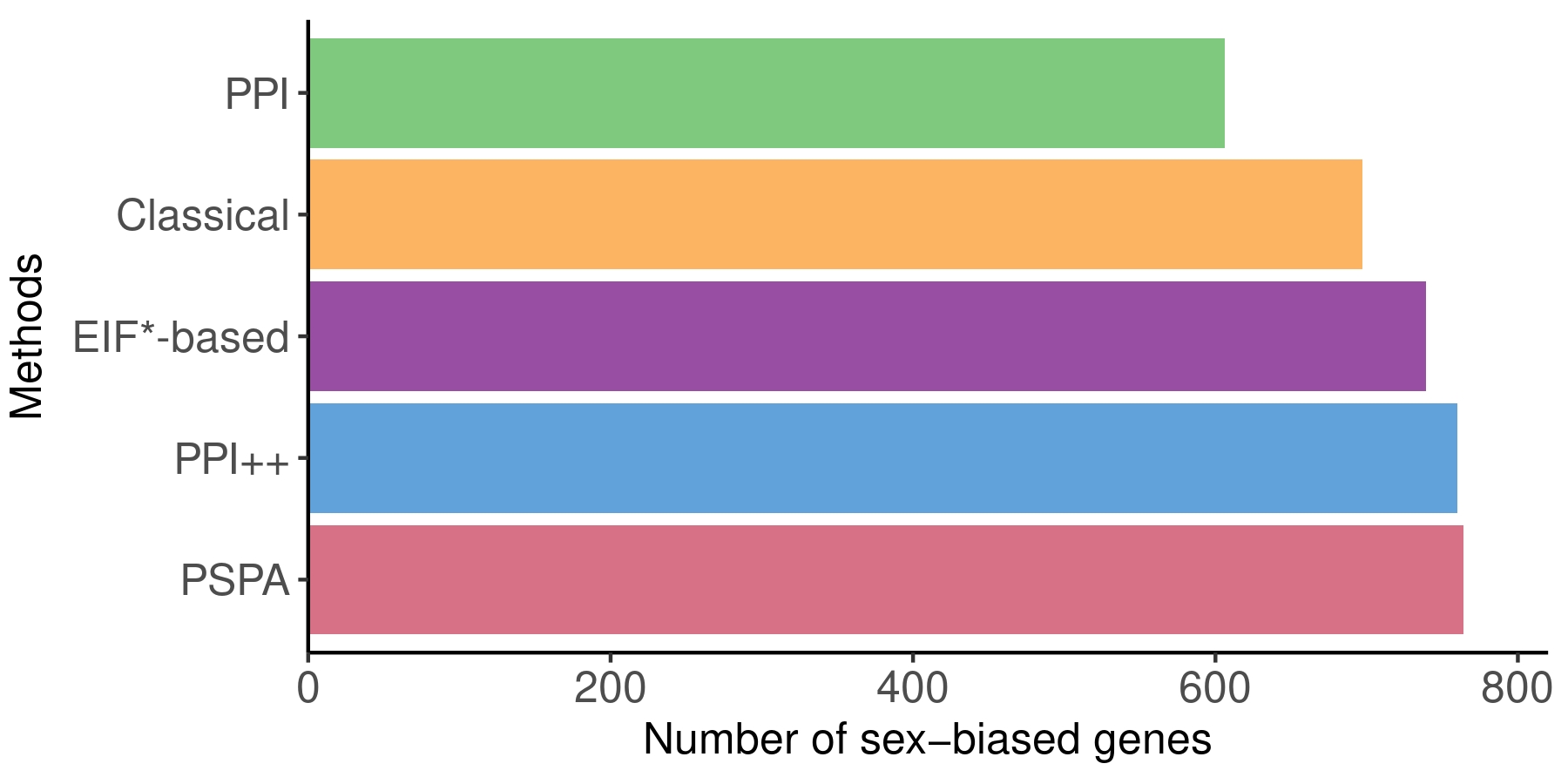}
    \caption{Comparison of \PPIpp~and \PSPA~in identifying sex-biased genes in GTEx.}
    \label{fig:GTex_PPI++}
\end{figure}

\begin{table}[H]
    \centering
    \renewcommand{\arraystretch}{1.25}
    \setlength{\tabcolsep}{3pt}
    \caption{\textbf{Comparison of \PPIpp~and \PSPA~in identifying risk factors for bone mineral density.} Estimate and $\wh{\text{SE}}$ represent the estimated linear regression coefficient and its corresponding standard error for each covariate. $\wh{\text{SE}}$ ratio indicates the ratio of the standard error for a specific method compared with the classical method. PA denotes physical activity, and SB refers to sedentary behavior. The bold font represents the method that gives the smallest $\wh{\text{SE}}$ ratio for each covariate.}
    \label{table:BMD_PPI++}
    \begin{tabular}{c|c|c|c|c|c|c|c}
        & & Biological sex & Age & PA & SB & Smoking & Alcohol \\ \hline
         \multirow{3}{*}{PPI++} & Estimate & -0.613 & -0.193 & 0.016 & 0.041 & -0.002 & -0.006 \\ \cline{2-8}
        & $\wh{\text{SE}}$ & 3.86E-03 & 3.79E-03 & 3.84E-03 & 3.91E-03 & 3.87E-03 & 3.89E-03 \\ \cline{2-8}
        & $\wh{\text{SE}}$ ratio & 0.920 & 0.919 & 0.943 & 0.935 & 0.939 & 0.926 \\ \hline
        \multirow{3}{*}{\PSPA} & Estimate & -0.614 & -0.181 & 0.010 & 0.040 & -0.008 & -5.15E-07 \\ \cline{2-8}
        & $\wh{\text{SE}}$ & 3.74E-03 & 3.74E-03 & 3.74E-03 & 3.82E-03 & 3.80E-03 & 3.81E-03 \\ \cline{2-8}
        & $\wh{\text{SE}}$ ratio & \textbf{0.892} & \textbf{0.907} & \textbf{0.919} & \textbf{0.914} & \textbf{0.920} & \textbf{0.908} \\ 
    \end{tabular}
\end{table}

\section{Proofs}
\subsection{Proof of Theorem \ref{thm:asy.normal.pspa}}
\label{proof:thm.asy.normal}

We first present a lemma that will be used later.
\begin{lemma}\label{lemma:unif.conv}
    Suppose that $X_1, \ldots, X_n$ are i.i.d., and the parameter space $\Theta$ is compact and $L(X, \theta)$ is continuous in $\theta \in \Theta$ almost everywhere. Moreover, there exists a function $h(X)$ satisfying $\|L(X, \theta)\| \leq h(X)$ for arbitrary $\theta \in \Theta$ and $\mathbb{E}\{h(X)\}<\infty$, then $\mathbb{E}\{L(X, \theta)\}$ is continuous in $\theta$ and
    \begin{align}
    \sup _{\theta \in \Theta}\left\|\frac{1}{n} \sum_{i=1}^n L\left(X_i, \theta\right)-\mathbb{E}\{L(X, \theta)\}\right\| \stackrel{P}{\rightarrow} 0.
    \end{align}
\end{lemma}
Lemma \ref{lemma:unif.conv} follows from \citet[Lemma 1]{tauchen1985diagnostic} and hence the proof is omitted.

Condition (C5): There exists a small neighborhood of $\bt$, denoted as $B$, in which $\bpsi(y,\x,\bt)$ is  continuously differentiable with respect to $\bt$ almost everywhere and the partial derivative satisfies $\E\sup_{\bt\in B}\lVert\partial\bpsi(Y,\X,\bt)/\partial\bt\rVert < \infty$ and $\E\sup_{\bt\in B}\lVert\partial\bpsi(\wh f(\Z),\X,\bt)/\partial\bt\rVert < \infty$.

For simplicity of notation, we re-write the estimating equation (\ref{eq:pspa.estimator}) as 
\begin{align}
      &  \Psi(\wh\bt\prop,\bo) =\Psi_n^Y + \diag(\bo)(-\Psi_n^{\wh f} + \Psi_N^{\wh f})= \0 , \text{ where}\\
       &\Psi_n^Y = \meann\psi(y_i,\x_i;\wh\bt\prop), \Psi_n^{\wh f} = \meann\psi(\wh f,\x_i;\wh\bt\prop), \text{ and }
       \Psi_N^{\wh f} = \meanN
       \psi(\wh f,\x_i;\wh\bt\prop).
    \end{align}
We first establish consistency following \citet[Theorem 5.9]{van2000asymptotic} by checking its two conditions. The deterministic condition is implied by Condition (C1) and (C2). We are left to verify the uniform convergence condition $\| \Psi(\wh\bt\prop) - \E(\Psi_n^Y)\|  = o_p(1)$.
By Condition (C3) and Lemma \ref{lemma:unif.conv}, we have 
\begin{align}
    \| \diag(\bo)\Psi_n^{\wh f} -\E(\diag(\bo)\Psi_n^{\wh f})\| &= o_p(1),\\
     \|\diag(\bo)\Psi_N^{\wh f}-\E(\diag(\bo)\Psi_N^{\wh f})\|   &= o_p(1),\text{ and}\\
     \|\Psi_n^Y-\E(\Psi_n^Y)\| &= o_p(1).
\end{align}
Hence, by triangular inequality,
\begin{align}
   & \| \Psi(\wh\bt\prop) - \E(\Psi_n^Y)\| \\
    \leq&\| \Psi(\wh\bt\prop) - \Psi_n^Y\| + \|\Psi_n^Y - \E(\Psi_n^Y)\|\\
    =& \|-\diag(\bo)\Psi_n^{\wh f} + \diag(\bo)\Psi_N^{\wh f}\| + 
        \| \Psi_n^Y - \E(\Psi_n^Y)\|\\
    \leq&\| -\bo\cdot\Psi_n^{\wh f} + \E(\diag(\bo)\Psi_n^{\wh f})\| + 
    \|\diag(\bo)\Psi_N^{\wh f}-\E(\diag(\bo)\Psi_N^{\wh f})\|  \\
    &
    + \|\E(\diag(\bo)\Psi_N^{\wh f} - \diag(\bo)\Psi_n^{\wh f})\|
    + \|\Psi_n^Y-\E(\Psi_n^Y)\|\\
    =& o_p(1),
\end{align}
which completes the proof of consistency, i.e., $\wh\bt\prop\convp\bt$. 
Next, we establish the asymptotic normality. Expanding $\Psi(\wh\bt\prop,\bo)$ at the true value $\bt$,
\begin{align}
    \0 =& \Psi(\bt,\bo) + (\wh\bt\prop-\bt)\partial\Psi(\Bar{\bt})/\partial\bt
\end{align}
 for some $\Bar{\bt}$ between $\bt$ and $\wh\bt\prop$. Multiplying both sides with $\sqrt n$ and with a direct calculation,
\begin{align}
 \left\{\partial\Psi(\Bar \bt)/\partial\bt\right\}\sqrt n (\wh\bt\prop-\bt) =& \sqrt n\Psi(\bt,\bo) .  
\end{align}
In order to show the asymptotic normality, by Slutsky's theorem, it's sufficient to verify the following:
\begin{align}
\label{eq:proof.asy.1}
    \|\partial\Psi(\Bar{\bt},\bo)/\partial\bt - \E\{\partial\bpsi(\bt)/\partial\bt\}\| = o_p(1),\\
\label{eq:proof.asy.2}
    \sqrt n\Psi(\bt,\bo) \coD \mathcal{N}(\0,\V(\bo)).
\end{align}
To show (\ref{eq:proof.asy.1}), we note that within the neighborhood $B$ in Condition (C5),
\begin{align}
    &\left\lVert\dfrac{\partial\Psi(\bt,\bo)}{\partial\bt} - \dfrac{\partial\Psi_n^Y(\bt)}{\partial\bt}\right\rVert\\
=& \left\lVert \diag(\bo)\left(-\partial\Psi_n^{\wh f}/\partial\bt + 
    \E\left[\partial\Psi_n^{\wh f}/\partial\bt\right]
    \right) +
    \diag(\bo)\left(\partial\Psi_N^{\wh f}/\partial\bt -
    \E\left[\partial\Psi_N^{\wh f}/\partial\bt\right]
    \right)
    \right\rVert\\
\leq & \left\lVert \diag(\bo)\left(-\partial\Psi_n^{\wh f}/\partial\bt + 
    \E\left[\partial\Psi_n^{\wh f}/\partial\bt\right]
    \right) \right\rVert +
    \left\lVert
    \diag(\bo)\left(\partial\Psi_N^{\wh f}/\partial\bt -
    \E\left[\partial\Psi_N^{\wh f}/\partial\bt\right]
    \right)
    \right\rVert.
\end{align}
Then together with Condition (C4) and Lemma \ref{lemma:unif.conv}, we have
\begin{align}\label{eq:proof.asy.1.to1}
    \sup_{\wt\bt\in B} \left\lVert\partial\Psi(\wt\bt)/\partial\wt\bt - \partial\Psi_n^Y(\wt\bt)/\partial\wt\bt\right\rVert
    &= o_p(1).
\end{align}
By Conditions (C1, (C4) and Lemma \ref{lemma:unif.conv}, we have 
\begin{align}\label{eq:proof.asy.1.to2}
     \sup_{\wt\bt\in B} \left\lVert\partial\Psi_n^Y(\wt\bt)/\partial\wt\bt - \E\left[\partial\Psi_n^Y(\wt\bt)/\partial\wt\bt\right]\right\rVert
    &= o_p(1).
\end{align}
Equations (\ref{eq:proof.asy.1.to1}) and (\ref{eq:proof.asy.1.to2}) give $\sup_{\wt\bt\in B} \left\lVert\partial\Psi(\wt\bt)/\partial\wt\bt - \E\left[\partial\Psi_n^Y(\wt\bt)/\partial\wt\bt\right]\right\rVert
    = o_p(1) $. Then the consistency of $\wh\bt\prop$ and the continuous mapping theorem imply that 
\begin{align}
 &\left\lVert\partial\Psi(\Bar{\bt},\bo)/\partial\bt - \E\left[\partial\bpsi(\bt)/\partial\bt\right] \right\rVert\\
 \leq &  
    \sup_{\wt\bt\in B} \left\lVert\partial\Psi(\wt\bt)/\partial\wt\bt - \E\left[\partial\Psi_n^Y(\bt)/\partial\bt\right]\right\rVert +
    \lVert \E\left[\partial\bpsi(\Bar{\bt})/\partial\bt\right] - 
             \E\left[\partial\bpsi(\bt)/\partial\bt\right]
    \rVert\\
\leq&o_p(1).
\end{align}
Therefore, condition (\ref{eq:proof.asy.1}) has been verified.
Meanwhile, (\ref{eq:proof.asy.2}) holds by Central Limit Theorem. The proof of asymptotic normality is completed.

\subsection{Proof of Corollary \ref{cor:asy.normal.est.w}}
\label{proof:asy.normal.est.w}
We denote the corresponding estimator as $\wh\bt(\wh\bo)$. Then the consistency of $\wh\bt(\wh\bo)$ follows from the consistency of $\wh\bo$ and the proof in Section \ref{proof:thm.asy.normal}. For asymptotic normality of $\wh\bt(\wh\bo)$, applying Taylor expansion on $0=\Psi(\wh\bt,\wh\bo)$ yields
\begin{align}
    0 &= \Psi(\wh\bt,\wh\bo)=\Psi(\bt,\bo^{\textnormal{opt}}) + \diag(\wh\bo - \bo^{\textnormal{opt}})
    \left(-\Psi_n^{\wh f} + \Psi_N^{\wh f}\right) + (\wh\bt-\bt)\partial\Psi(\Bar{\bt})/\partial\bt
\end{align}
for some $\Bar{\bt}$ between $\bt$ and $\wh\bt(\wh\bo)$. With the same proof technique in Section \ref{proof:thm.asy.normal}, we have 
\begin{align}
    \sqrt n(\wh\bt(\wh\bo)-\bt)\coD N(\0,\bS(\bo^{\textnormal{opt}})),
\end{align}
which implies Corollary \ref{cor:asy.normal.est.w}.

\subsection{Proof of Proposition \ref{prop:opt}}
\label{proof:prop:opt}
The $j$-th diagonal element of the asymptotic covariance matrix $\bS(\bo)$ of $ \wh\bt\prop({\bo})$ given weighting vector $\bo$ is
\begin{align}
    \bS_{jj}(\bo) 
    =
    \omega_j^2[\A\inv(\M_2+\rho\M_3)\A\inv]_{jj} - 2\omega_j[\A\inv\M_4\A\inv]_{jj} + [\A\inv\M_1\A\inv]_{jj}.
\end{align}

It is a quadratic function of $\omega_j$ with a unique minimizer
\begin{align}
    \omega_j^{\textnormal{opt}} = \dfrac{[\A\inv\M_4\A\inv]_{jj}}{[\A\inv(\M_2+\rho\M_3)\A\inv]_{jj}}.
\end{align}

Therefore, given any weighting vector $\bo^* = [\omega_1^*, \cdots, \omega_q^*]\trans$, $\bS_{jj}(\bo^{\textnormal{opt}}) \leq \bS_{jj}(\bo^*)$, which completes the proof.

\subsection{Proof of Proposition \ref{prop:eif}}\label{proof:prop.eif}
The joint log-likelihood for a generic observation $(r,\x,\z,y)$ is 
\begin{align}
l &= r\log p (y\mid\x,\z) + \log p(\x\mid\z) + \log p(\z) + 
r\log(\pi) + (1-r)\log(1-\pi).
\end{align}
Taking the semiparametric approach \citep{bickel1993efficient,tsiatis2006semiparametric}, we consider the Hilbert space $\calH$ that contains all one-dimensional zero-mean measurable functions of the observed data with finite variance, equipped with the inner product $\langle h_1, h_2\rangle = \E\{h_1(\cdot) h_2(\cdot)\}$ where $h_1, h_2\in \calH$.
To estimate $\bt$, we regard  $p(y\mid\x,\z)$, $p(\x\mid\z)$ and $p(\z)$ as nuisance functions.
Denote their nuisance tangent spaces by $\calT_y$, $\calT_{\x}$, and $\calT_{\z}$, which are defined as the mean squared closure of the nuisance tangent spaces of parametric submodels spanned by the nuisance score vectors.
We have that $\calT  =\calT_y\oplus\calT_{\x}\oplus\calT_{\z}$ where
\begin{align}
\calT_y &= [r\a_1(y,\x,\z): \E\{\a_1(Y,\x,\z)\mid\x,\z\} = \0],\\
\calT_{\x} &= [\a_2(\x,\z) :\E\{\a_2(\X,\z)\mid\z\} = \0],\text{ and}\\
\calT_{\z} &= [\a_3(\z),\E\{\a_3(\Z)\} = \0],
\end{align}
respectively, and the notation $\oplus$ represents the direct sum of two spaces that are orthogonal to each other.

We introduce parametric submodels $p_{\bta_y}(y\mid\x,\z)$, $p_{\bta_{\x}} (\x\mid\z)$,  and $p_{\bta_{\z}}(\z)$, where $\bta = [\bta_y\trans,\bta_{\x}\trans,\bta_{\z}\trans,\bta_r\trans]\trans$ is a vector of nuisance parameters. Let 
\begin{align}
\S_{\bta_y}(y,\x,\z) = \dfrac{\partial\log p_{\bta_y}(y,\x,\z)}{\partial\bta_y},\ 
\S_{\bta_{\x}}(\x,\z) = \dfrac{\partial\log p_{\bta_{\x}}(\x,\z)}{\partial\bta_{\x}},\text{ and }
\S_{\bta_{\z}}(\z) = \dfrac{\partial\log p_{\bta_{\z}}(\z)}{\partial\bta_{\z}},
\end{align}
then these scores functions satisfy
$r\S_{\bta_y}(y,\x,\z) \in \calT_y,\ 
\S_{\bta_{\x}} (\x,\z)\in \calT_{\x},\text{ and }
\S_{\bta_{\z}} (\z) \in \calT_{\z}$.
Recall the definition of parameter of interest, i.e.,
\begin{align}
\iiint \bpsi(y,\x;\bt) p(y\mid\x,\z)p(\x\mid\z)p(\z)\dd \z\dd\x\dd y = \0.
\end{align}
Taking derivative of $\bt$ with respect to nuisance parameters,
\begin{align}
\dfrac{\partial\bt}{\partial\bta_y\trans} =& \A\inv\dfrac{\partial\E\{\bpsi(Y,\X;\bt)\}}{\partial\bta_y\trans} = \A\inv\E[\bpsi(Y,\X;\bt)\S_{\bta_y}\trans(Y,\X,\Z)],\\
\dfrac{\partial\bt}{\partial\bta_{\x}\trans} =& \A\inv\dfrac{\partial\E\{\bpsi(Y,\X;\bt)\}}{\partial\bta_{\x}\trans} = \A\inv\E\{\bpsi(Y,\X;\bt)\S_{\bta_{\x}}\trans(\X,\Z)\},\text{ and}\\
\dfrac{\partial\bt}{\partial\bta_{\z}\trans} =& \A\inv\dfrac{\partial\E\{\bpsi(Y,\X;\bt)\}}{\partial\bta_{\z}\trans} = \A\inv\E\{\bpsi(Y,\X;\bt)\S_{\bta_{\z}}\trans(\Z)\}.
\end{align}
Let $\bphi(y,\x;\bt) = r\a_1(y,\x,\z) + \a_2(\x,\z)+ \a_3(\z)$, where
\begin{align}
\a_1(y,\x,\z) = & \dfrac{1}{\pi(\z)} \A\inv\left[\bpsi(y,\x;\bt) - \E\{\bpsi(Y,\x;\bt)\mid\x,\z\}\right],\\
\a_2(\x,\z) =& \A\inv \left[
    \E\{\bpsi(Y,\x;\bt)\mid\x,\z\} - \E\{\bpsi(Y,\x;\bt)\mid\x,\z\}
    \right],\text{ and}\\
\a_3(\z) =& \A\inv\E\{\bpsi(Y,\x;\bt)\mid\x,\z\}.
\end{align}
Then it can be easily verified that $r\a_1(y,\x,\z)\in\calT_y$, $\a_2(\x,\z)\in\calT_{\x}$, $\a_3(\z)\in\calT_{\x}$, and 
\begin{align}
\small
    \E\{\bphi(Y,\X;\bt) R\S_{\bta_y}\trans(Y,\X,\Z)\} = \dfrac{\partial\bt}{\partial\bta_y\trans},\ 
    \E\left\{\bphi(Y,\X;\bt)\S_{\bta_{\x}}\trans(\X,\Z))\right\} = \dfrac{\partial\bt}{\partial\bta_{\x}\trans},\text{ and }
    \E\{\bphi(Y,\X;\bt)\S_{\bta_{\z}}\trans(\Z)\} = \dfrac{\partial\bt}{\partial\bta_{\z}\trans}.
\end{align}
Therefore, $\bphi(y,\x;\bt)$ is the efficient influence function. The proof of Proposition \ref{prop:eif} is completed.

\subsection{Lemma for \PSPA~when applied to ML-predicted covariates}
\begin{lemma}\label{lem:eigen}
    Letting $\A$ and $\B$ be two $q \times q$ gram matrices, and $\bo$ is a q-dimensional vector. If $0 \leq \omega_j \leq \dfrac{\lambda_{{\min, +}}(\A)}{\lambda_{\max}(\B)}$ for all $j \in \{1, \cdots, p\}$
    then $\A - \text{diag}(\bo)\B$ is positive semi-definite.
\end{lemma}

To prove the lemma, we will verify that $\A-\operatorname{diag}(\bo) \B$ is positive semi-definite by showing that for any vector $\x$, the quadratic form $\x^T(\A-\operatorname{diag}(\bo) \B) \x \geq 0$.

Let $\x$ be any vector in $\mathbb{R}^p$. Since
\begin{align}
    \x\trans \A x &\geq \lambda_{\min, +} (\A)\|\x\|^2 \\
    \x\trans\operatorname{diag}(\bo) \B \x &\leq \lambda_{\max} (\operatorname{diag}(\bo) \B)\|x\|^2 \leq \max_j \omega_j\lambda_{\max}\B\|\x\|^2
\end{align}

Therefore
\begin{align}
    \x\trans \A \x - \x\trans\operatorname{diag}(\bo) \B \geq 
    (\lambda_{\min, +} (\A) - \max_j \omega_j\lambda_{\max}\B)\|\x\|^2
\end{align}

To ensure $\x\trans \A \x - \x\trans\operatorname{diag}(\bo) \B \geq  0$, we require $ \lambda_{\min, +} (\A) - \max_j \omega_j\lambda_{\max}(\B) \geq 0$, that is
\begin{align}
    \max_j \omega_j \leq \dfrac{ \lambda_{\min, +} (\A)}{\lambda_{\max}(\B)}
\end{align}

\subsection{Proof of Corollary \ref{prop:opt:x}}
\label{proof:prop:opt:x}
The $j$-th diagonal element of the asymptotic covariance matrix $\bS'(\bo)$ of $\wh\bt_{\PSPA'}$ given weighting vector $\bo$ is
\begin{align}
    \bS'_{jj}(\bo) 
    &=
    \omega_j^2[\A\inv(\M_2'+\rho\M_3')\A\inv]_{jj} - 2\omega_j[\A\inv\M_4'\A\inv]_{jj} + [\A\inv\M_1'\A\inv]_{jj},
\end{align}

It is a quadratic function of $\omega_j$ with a unique minimizer
\begin{align}
    \omega_j^{\textnormal{opt}, *} = \dfrac{[\A\inv\M'_4\A\inv]_{jj}}{[\A\inv(\M'_2+\rho\M'_3)\A\inv]_{jj}}.
\end{align}

If $\omega_j^{\textnormal{opt}, *} >0$, $\bS'_{jj}(\bo)$ is a decreasing function with $\omega_j \in [0, \omega_j^{\textnormal{opt}, *}]$. If $\omega_j^{\textnormal{opt}, *} \leq 0$, $\bS'_{jj}(\bo)$ $\bS'_{jj}(\bo)$ is an increasing function with $\omega_j \in [\omega_j^{\textnormal{opt}, *}, 0]$. We also have $\frac{\lambda_{{\min, +}}[\frac{1}{n}\nabla\bpsi(y_i,\x_i;\bt)]}{\lambda_{\max}[\frac{1}{n}\nabla\bpsi(y_i,\wh q;\bt)]} >0$

Therefore, given $\hat\bo^{\textnormal{opt}} = [\wh\omega_1^{\textnormal{opt}},\dots,\wh\omega_q^{\textnormal{opt}}]\trans$, where 
\begin{small}
\begin{align}
\wh\omega_j^{\textnormal{opt}} =
\begin{cases}
\min(\frac{[\wh\A\inv\wh\M_{4,{\text{C}}}^{'}\wh\A\inv]_{jj}}{[\wh\A\inv(\wh\M_{2,{\text{C}}}^{'}+\rho\wh\M_{3,{\text{C}}}^{'})\wh\A\inv]_{jj}}, \frac{\lambda_{{\min, +}}[\frac{1}{n}\nabla\bpsi(y_i,\x_i;\bt)]}{\lambda_{\max}[\frac{1}{n}\nabla\bpsi(y_i,\wh q;\bt)]}) &\quad\text{if } \frac{[\wh\A\inv\wh\M_{4,{\text{C}}}^{'}\wh\A\inv]_{jj}}{[\wh\A\inv(\wh\M_{2,{\text{C}}}^{'}+\rho\wh\M_{3,{\text{C}}}^{'})\wh\A\inv]_{jj}} >0  \\
\max(\frac{[\wh\A\inv\wh\M_{4,{\text{C}}}^{'}\wh\A\inv]_{jj}}{[\wh\A\inv(\wh\M_{2,{\text{C}}}^{'}+\rho\wh\M_{3,{\text{C}}}^{'})\wh\A\inv]_{jj}}, 0) &\quad\text{if } \frac{[\wh\A\inv\wh\M_{4,{\text{C}}}^{'}\wh\A\inv]_{jj}}{[\wh\A\inv(\wh\M_{2,{\text{C}}}^{'}+\rho\wh\M_{3,{\text{C}}}^{'})\wh\A\inv]_{jj}} \leq 0  \\
\end{cases}
\end{align}
\end{small}
for all $j \in 1, \dots, q$, and the weighting vector $\bo^{\textnormal{C}} = [0, \cdots, 0]\trans$ that correspond to the classical estimator, $\bS_{jj}(\bo^{\textnormal{opt}}) \leq \bS_{jj}(\bo^{\textnormal{C}})$, which completes the proof.

\bibliographystyle{apalike}
\bibliography{main.bib}

\begin{thebibliography}{}

\bibitem[Angelopoulos et~al., 2023a]{angelopoulos2023prediction}
Angelopoulos, A.~N., Bates, S., Fannjiang, C., Jordan, M.~I., and Zrnic, T. (2023a).
\newblock Prediction-powered inference.
\newblock {\em Science}, 382(6671):669--674.

\bibitem[Angelopoulos et~al., 2023b]{angelopoulos2023ppi++}
Angelopoulos, A.~N., Duchi, J.~C., and Zrnic, T. (2023b).
\newblock Ppi++: Efficient prediction-powered inference.
\newblock {\em arXiv preprint arXiv:2311.01453}.

\bibitem[Azriel et~al., 2022]{azriel2022semi}
Azriel, D., Brown, L.~D., Sklar, M., Berk, R., Buja, A., and Zhao, L. (2022).
\newblock Semi-supervised linear regression.
\newblock {\em Journal of the American Statistical Association}, 117(540):2238--2251.

\bibitem[Basu et~al., 2021]{basu2021predicting}
Basu, M., Wang, K., Ruppin, E., and Hannenhalli, S. (2021).
\newblock Predicting tissue-specific gene expression from whole blood transcriptome.
\newblock {\em Science Advances}, 7(14):eabd6991.

\bibitem[Bickel et~al., 1993]{bickel1993efficient}
Bickel, P.~J., Klaassen, C.~A., Bickel, P.~J., Ritov, Y., Klaassen, J., Wellner, J.~A., and Ritov, Y. (1993).
\newblock {\em Efficient and adaptive estimation for semiparametric models}, volume~4.
\newblock Springer.

\bibitem[Breidt and Opsomer, 2017]{breidt2017model}
Breidt, F.~J. and Opsomer, J.~D. (2017).
\newblock Model-assisted survey estimation with modern prediction techniques.

\bibitem[Bullock et~al., 2020]{bullock2020satellite}
Bullock, E.~L., Woodcock, C.~E., Souza~Jr, C., and Olofsson, P. (2020).
\newblock Satellite-based estimates reveal widespread forest degradation in the amazon.
\newblock {\em Global Change Biology}, 26(5):2956--2969.

\bibitem[Chakrabortty and Cai, 2018]{chakrabortty2018efficient}
Chakrabortty, A. and Cai, T. (2018).
\newblock Efficient and adaptive linear regression in semi-supervised settings.

\bibitem[Chen et~al., 2005]{chen2005measurement}
Chen, X., Hong, H., and Tamer, E. (2005).
\newblock Measurement error models with auxiliary data.
\newblock {\em The Review of Economic Studies}, 72(2):343--366.

\bibitem[{GTEx Consortium}, 2020]{gtex2020gtex}
{GTEx Consortium} (2020).
\newblock The gtex consortium atlas of genetic regulatory effects across human tissues.
\newblock {\em Science}, 369(6509):1318--1330.

\bibitem[{GTEx Consortium} et~al., 2015]{gtex2015genotype}
{GTEx Consortium}, Ardlie, K.~G., Deluca, D.~S., Segr{\`e}, A.~V., Sullivan, T.~J., Young, T.~R., Gelfand, E.~T., Trowbridge, C.~A., Maller, J.~B., Tukiainen, T., et~al. (2015).
\newblock The genotype-tissue expression (gtex) pilot analysis: multitissue gene regulation in humans.
\newblock {\em Science}, 348(6235):648--660.

\bibitem[Hahn, 1998]{hahn1998role}
Hahn, J. (1998).
\newblock On the role of the propensity score in efficient semiparametric estimation of average treatment effects.
\newblock {\em Econometrica}, pages 315--331.

\bibitem[Kallus and Mao, 2020]{kallus2020role}
Kallus, N. and Mao, X. (2020).
\newblock On the role of surrogates in the efficient estimation of treatment effects with limited outcome data.
\newblock {\em arXiv preprint arXiv:2003.12408}.

\bibitem[LeCun et~al., 2015]{lecun2015deep}
LeCun, Y., Bengio, Y., and Hinton, G. (2015).
\newblock Deep learning.
\newblock {\em nature}, 521(7553):436--444.

\bibitem[Lonsdale et~al., 2013]{lonsdale2013genotype}
Lonsdale, J., Thomas, J., Salvatore, M., Phillips, R., Lo, E., Shad, S., Hasz, R., Walters, G., Garcia, F., Young, N., et~al. (2013).
\newblock The genotype-tissue expression (gtex) project.
\newblock {\em Nature genetics}, 45(6):580--585.

\bibitem[Molenberghs et~al., 2014]{molenberghs2014handbook}
Molenberghs, G., Fitzmaurice, G., Kenward, M.~G., Tsiatis, A., and Verbeke, G. (2014).
\newblock {\em Handbook of missing data methodology}.
\newblock CRC Press.

\bibitem[Robins, 2000]{robins2000robust}
Robins, J.~M. (2000).
\newblock Robust estimation in sequentially ignorable missing data and causal inference models.
\newblock In {\em Proceedings of the American Statistical Association}, volume 1999, pages 6--10. Indianapolis, IN.

\bibitem[Robins et~al., 1994]{robins1994estimation}
Robins, J.~M., Rotnitzky, A., and Zhao, L.~P. (1994).
\newblock Estimation of regression coefficients when some regressors are not always observed.
\newblock {\em Journal of the American statistical Association}, 89(427):846--866.

\bibitem[Song et~al., 2023]{song2023general}
Song, S., Lin, Y., and Zhou, Y. (2023).
\newblock A general m-estimation theory in semi-supervised framework.
\newblock {\em Journal of the American Statistical Association}, pages 1--11.

\bibitem[Tauchen, 1985]{tauchen1985diagnostic}
Tauchen, G. (1985).
\newblock Diagnostic testing and evaluation of maximum likelihood models.
\newblock {\em Journal of Econometrics}, 30(1):415--443.

\bibitem[Tsiatis, 2006]{tsiatis2006semiparametric}
Tsiatis, A.~A. (2006).
\newblock Semiparametric theory and missing data.

\bibitem[Van~der Vaart, 2000]{van2000asymptotic}
Van~der Vaart, A.~W. (2000).
\newblock {\em Asymptotic statistics}, volume~3.
\newblock Cambridge university press.

\bibitem[Vi{\~n}as et~al., 2023]{vinas2023hypergraph}
Vi{\~n}as, R., Joshi, C.~K., Georgiev, D., Lin, P., Dumitrascu, B., Gamazon, E.~R., and Li{\`o}, P. (2023).
\newblock Hypergraph factorization for multi-tissue gene expression imputation.
\newblock {\em Nature machine intelligence}, 5(7):739--753.

\bibitem[Wang et~al., 2023]{wang2023scientific}
Wang, H., Fu, T., Du, Y., Gao, W., Huang, K., Liu, Z., Chandak, P., Liu, S., Van~Katwyk, P., Deac, A., et~al. (2023).
\newblock Scientific discovery in the age of artificial intelligence.
\newblock {\em Nature}, 620(7972):47--60.

\bibitem[Wang et~al., 2016]{wang2016imputing}
Wang, J., Gamazon, E.~R., Pierce, B.~L., Stranger, B.~E., Im, H.~K., Gibbons, R.~D., Cox, N.~J., Nicolae, D.~L., and Chen, L.~S. (2016).
\newblock Imputing gene expression in uncollected tissues within and beyond gtex.
\newblock {\em The American Journal of Human Genetics}, 98(4):697--708.

\bibitem[Wang and Shen, 2007]{wang2007large}
Wang, J. and Shen, X. (2007).
\newblock Large margin semi-supervised learning.
\newblock {\em Journal of Machine Learning Research}, 8(8).

\bibitem[Wang et~al., 2020]{wang2020methods}
Wang, S., McCormick, T.~H., and Leek, J.~T. (2020).
\newblock Methods for correcting inference based on outcomes predicted by machine learning.
\newblock {\em Proceedings of the National Academy of Sciences}, 117(48):30266--30275.

\bibitem[Zhang et~al., 2019]{zhang2019semi}
Zhang, A., Brown, L.~D., and Cai, T.~T. (2019).
\newblock Semi-supervised inference: General theory and estimation of means.

\end{thebibliography}

\end{document}